\documentclass{ar-1col-S2O}

\newcommand*\araa{ARA\&A}

\usepackage[comma]{natbib}
\usepackage{url}
\usepackage{anyfontsize}
\usepackage{amssymb}

\def\jnlref#1{{\rm#1}}
\def\actaa{\jnlref{ACTA}}
\def\aj{\jnlref{AJ}}
\def\araa{\jnlref{ARA\&A}}
\def\apj{\jnlref{ApJ}}
\def\apjl{\jnlref{ApJ}}

\def\apjs{\jnlref{ApJS}}

\def\apss{\jnlref{Ap\&SS}}
\def\aap{\jnlref{A\&A}}
\def\aapr{\jnlref{A\&A~Rev.}}

\def\mnras{\jnlref{MNRAS}}
\def\nar{\jnlref{NewAR}}
\def\na{\jnlref{NewA}}
\def\nat{\jnlref{Nature}}

\def\prd{\jnlref{Phys.~Rev.~D}}

\def\prl{\jnlref{Phys.~Rev.~Lett.}}
\def\pasa{\jnlref{PASA}}
\def\pasp{\jnlref{PASP}}

\def\zap{\jnlref{ZAp}}

\def\bain{\jnlref{Bull.~Astron.~Inst.~Netherlands}}

\def\physrep{\jnlref{Phys.~Rep.}}

\jname{Xxxx. Xxx. Xxx. Xxx.}
\jvol{AA}
\jyear{2023}
\doi{10.1146/((please add article doi))}

\begin{document}

\markboth{Marchant \& Bodensteiner}{Massive Binary Stars}

\title{The Evolution of Massive Binary Stars}

\author{Pablo Marchant,$^{1}$ and Julia Bodensteiner$^{2}$
\affil{$^{1}$Institute of Astronomy, KU Leuven, Celestijnlaan 200D, 3001 Leuven, Belgium; email: pablo.marchant@kuleuven.be.}
\affil{$^{2}$ESO -- European Organisation for Astronomical Research in the Southern Hemisphere, Karl-Schwarzschild-Strasse 2, 85748 Garching, Germany; email: julia.bodensteiner@eso.org.}}

\begin{abstract}
  Massive stars play a major role in the evolution of their host galaxies, and serve as important probes of the distant Universe. It has been established that the majority of massive stars reside in close binaries and will interact with their companion stars during their lifetime. Such interactions drastically alter their life cycles and complicate our understanding of their evolution, but are also responsible for the production of interesting and exotic interaction products.

  \begin{itemize}
    \item Extensive observation campaigns with well-understood detection
    
    sensitivities have allowed to convert the observed properties into
    
    intrinsic characteristics, facilitating a direct comparison to theory.
    \item Studies of large samples of massive stars in our Galaxy and the
    
    Magellanic Clouds have unveiled new types of interaction products,
    
    providing critical constraints on the mass transfer phase and the
    
    formation of compact objects.
    \item The direct detection of gravitational waves has revolutionized
    
    the study of stellar mass compact objects, providing a new window
    
    to study massive star evolution. Their formation processes are,
    
    however, still unclear. The known sample of compact object
    
    mergers will grow by orders of magnitude in the coming decade,
    
    turning into the best understood astrophysical population.
  \end{itemize}
\end{abstract}

\begin{keywords}
massive stars, binary star, stellar evolution, rotation, compact objects, gravitational waves
\end{keywords}

\maketitle

\tableofcontents

\section{INTRODUCTION}\label{sec:intro}

Massive stars are powerful cosmic engines, capable of modifying their local, galactic and even extragalactic environments \citep{Hopkins+2014}. Through their high luminosities, which can include a significant fraction of ionizing radiation, they are understood to play a critical role in the re-ionization of the Universe \citep{HaimanLoeb1997}. Radiation driven stellar winds can remove large fractions of the stellar birth mass \citep{Vink2022}, providing kinetic feedback and chemically processed matter. At the end of their lives, energetic supernova explosions (SNe) further enrich their surroundings with matter that has undergone late nuclear burning stages \citep{Nomoto+2013}, acting as one of the main drivers of the chemical evolution of galaxies. Although rare objects, massive stars dominate the integrated light of distant galaxies \citep[e.g.][]{Pettini2000}. Yet, despite their broad astrophysical importance, various physical processes that dominate their evolution are still largely uncertain (see \citealt{Langer2012} for a recent review).

One critical aspect that undermines the understanding of massive stars is their scarcity. In the local Universe $<1\%$ of stars born are expected to be massive \citep{Kroupa2002}, which is further compounded with their orders of magnitude shorter lifetime compared to intermediate and low-mass stars. But another critical complication arises from the prevalence of close binaries in massive stars. For several decades multi-epoch spectroscopic observations have indicated that a large fraction of massive stars have close binary companions \citep{Levato1987, Abt1990, Kobulnicky+2007}. By carefully accounting for observational biases it has been shown that the majority of massive stars undergo interaction phases that dominate their evolution, potentially with over half of them interacting before the end of the main-sequence \citep{Sana2012}. Detections made through photometry, astrometry or other techniques that cover different regimes not accessible by spectroscopy further indicate that a large number of stars have multiple companions (see \citealt{Moe2017} for a recent compilation). Binary evolution adds on the complexity of massive star physics, providing a rich set of post-interaction products such as stripped stars \citep{Shenar2020_LB1, Drout2023}, rapidly rotating accretors \citep{deMink+2013, Renzo2021}, mergers \citep{Schneider+2019, Hirai+2021}, exotic supernovae \citep{Chevalier2012, Metzger+2022}, X-ray binaries \citep{TaurisvandenHeuvel2006, Gilfanov+2022} and double-degenerate binaries which may produce detectable gravitational wave (GW) emission \citep{Tauris+2017,MandelFarmer2022}. Observations of apparently single stars can hide a past of binary interaction, leading to an erroneous attribution of their physical properties to single-star physics. 

Even if theory can describe the intrinsic properties of a given stellar population, comparison to observations requires a clear understanding of the biases involved in sample selection and instrumental limitations. Large-scale surveys with clearly defined selection criteria play a critical role in this regard. This is the case for spectroscopic surveys such as the VLT-FLAMES Tarantula Survey \citep[VFTS, see][for an overview]{Evans2011}, photometric surveys such as Kepler \citep{Borucki2010, Koch2010} and TESS \citep{Ricker2015}, and the astrometric measurements of the GAIA mission \citep{Gaia2016}. The observation of transients through multi-band synoptic surveys has also provided breakthroughs on our understanding of SNe. For instance, the Zwicky Transient Facility \citep[ZTF,][]{Bellm2019} currently produces on the order of a million daily transient alerts, with dedicated infrastructure to perform spectroscopic follow-up and classification of the brightest ones \citep{Fremling+2020}. Entire populations of high-mass X-ray binaries (HMXBs) can be probed in individual galaxies with X-ray observations \citep{Gilfanov+2022}. However, the most significant development in the past decade did not involve electromagnetism, but rather the direct detection of gravitational waves (GWs) from merging compact objects \citep{GW150914}. Ground-based interferometers such as LIGO \citep{LIGO2015} and Virgo \citep{VIRGO2015} have well-understood biases, making it straightforward to determine intrinsic population properties, or to apply biases synthetically to predicted populations. 

Owing to the large progress seen in the last decade, a review of the current state and standing problems in the field of massive binary evolution is of critical importance.
However, a full comprehensive review is not possible within the scope of this document, and as such, our focus is on the progress that has been made since the review of \citet{Langer2012} on single and binary massive-star evolution. Our scope is also limited to avoid overlap with the recent reviews of \cite{EldridgeStanway2022} on the impact of massive binaries on the evolution of early galaxies, \cite{Kaaret+2017} on ultraluminous X-ray sources and \cite{MuraseBartos2019} on multi-messenger astrophysics. We also exclude from the discussion the evolution of low-mass X-ray binaries, whose  
formation is potentially affected by stellar dynamics (eg. \citealt{Ivanova+2010}). The field of transients and supernovae associated to binary evolution is also rapidly growing, and cannot be comprehensively covered in this review (for a recent overview, see chapter 13 of \citealt{TaurisVandenHeuvel2023}). We do, however, soften the definition of 'massive star'\footnote{\citet{Langer2012} defined a massive star as ``a star that is  massive enough to form a collapsing core at the end of its life and, thus, avoid
the white dwarf fate''.} and include observations and findings for intermediate-mass stars when applicable to the massive star regime.

The outline of this review is as follows: In Section \ref{sec:binarity} we briefly mention the observational methods used to detect and characterize binaries, and discuss the constraints obtained on binary fractions and orbital parameters at different evolutionary stages. In Section \ref{sec:interaction} we review the main binary-interaction processes, and discuss the theoretical tools used to model them. We then discuss the properties of non-degenerate post-interaction products in Section \ref{sec:post}, and how their observation constrains our theoretical models. In Section \ref{sec:deg} we discuss the properties of single- and double-degenerate interaction products. In Section \ref{sec:GW} we provide a brief review on the observations of gravitational-wave sources, and their potential formation through binary evolution. Finally, we conclude in Section\,\ref{sec:concl}.

\section{CONSTRAINTS ON BINARITY AT DIFFERENT EVOLUTIONARY STAGES}\label{sec:binarity}

The consensus that massive stars are predominantly part of binary or higher-order multiple systems has been established over the last decades using a multitude of different observations and detection techniques. Those probe different regions of the parameter space and suffer from different observational biases and limitations. 

Photometric signatures such as eclipses and ellipsoidal variations can only be detected in the closest, shortest-period binaries, if their orientation towards the observer is favorable. 
Spectroscopic observations use radial velocity (RV) variations to detect binaries and constrain orbital parameters such as the period and eccentricity in the case of single-lined spectroscopic binaries (SB1s), where only one component, usually denoted as primary, is visible. Additionally, the mass ratio can be measured in double-lined spectroscopic binaries (SB2s), where the signature of both stars is discernible in the spectrum (it is usually defined as the mass of the less luminous star, the secondary, over the mass of the primary). Spectroscopic observations probe orbital periods that are of the order of the length of the observing campaign (usually up to a few years) but also suffer from severe observational biases depending on the system's parameters and orientation \citep[e.g.,][and see Section\,\ref{subsec:binstats_OB}]{Sana2012}. 

Long-baseline interferometry, which is still only technically feasible for brighter stars in our own Galaxy with current instrumentation, allows to probe binary systems that have angular separations between $\sim$1 and 100\,mas, corresponding to orbital periods of the order of months to decades for Galactic distances \citep[e.g.,][]{LeBouquin2017, Karl2018, Bordier2022, Lanthermann2023}. 
Similarly, high-precision long-term astrometric monitoring allows the detection of wider binaries. While barely any massive stars are included in the current Gaia data release 3 \citep[DR3][see also Section\,\ref{sec:inert}]{Gaia2021}, the fourth data release (DR4), expected towards the end of 2025, should improve this situation significantly and provide constraints on a large number of massive binaries. 
Binaries with even larger separations or more distant companions in higher-order multiples (with angular separations of $\sim$100 -- 10\,000\,mas) can further be detected with high-angular resolution imaging techniques such as high-contrast or AO-supported imaging \citep[e.g.,][]{Mason2009, Sana2014, Aldoretta2015, Kalari2022, Reggiani2022, Pauwels2023}.

\begin{figure}
    \centering
    \includegraphics[width=\columnwidth]{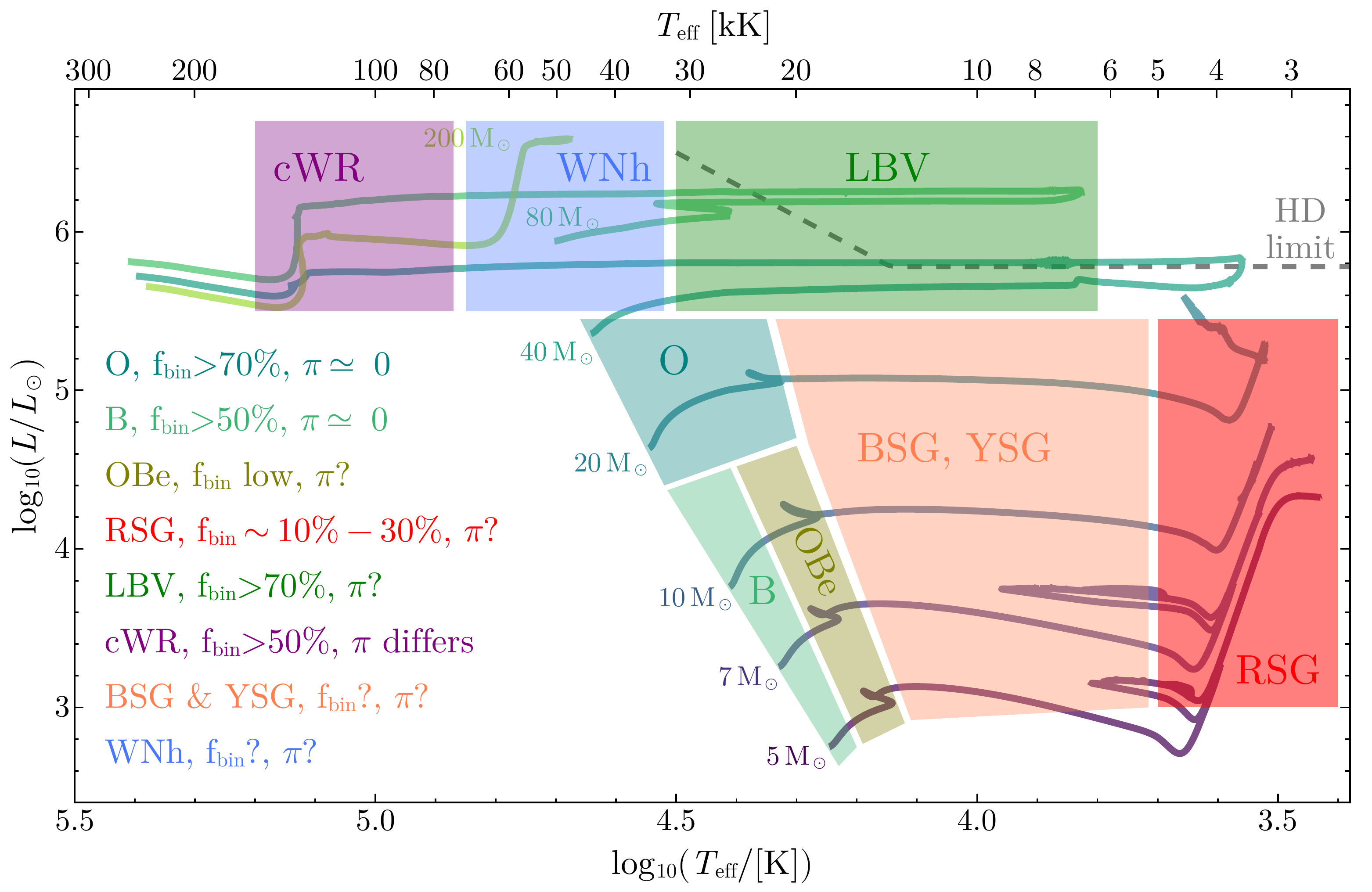}
    \caption{Binary fractions and the index of the period distribution $\pi$ for different classes of stars and evolutionary stages across the HRD. Stellar evolution tracks of different masses are shown, covering evolution from the zero-age MS until core-carbon depletion (or until just before the hydrogen envelope is removed in the AGB phase for the $5\;\mathrm{M_\odot}$ and $7\;\mathrm{M_\odot}$ models). The dashed line indicates the Humphreys-Davidson (HD) limit, above which there is a lack of observed stars \citep{HumphreysDavidson1979}. See text for references.}
    \label{fig:HRD_binary_stats}
\end{figure}

Given that binary interactions are thought to occur in close binaries with periods below $\sim10$\,years \citep[e.g.,][and Section\,\ref{sec:interaction}]{Podsiadlowski1992}, we focus on spectroscopic surveys in the following, and only briefly mention complementary observations using other techniques. Thereby we go through different evolutionary stages, from the main sequence (MS) to further evolved stars. An overview of all reported binary properties is displayed in Figure\,\ref{fig:HRD_binary_stats}, while references are provided in the text\footnote{All stellar evolution tracks shown in this review were computed using the \texttt{MESA} stellar evolution code at solar metallicity (Z=0.0142, \citealt{Asplund+2009}). All data needed to reproduce our figures and simulations is provided at \url{https://doi.org/10.5281/zenodo.10009234}}.

\subsection{OB stars}\label{subsec:binstats_OB}
Several spectroscopic studies targeting the multiplicity properties of OB stars at different metallicity were performed in recent years (for an overview, see Table\,\ref{tab:bin_stats}). As most interactions occur after the MS when the primary expands (see Section\,\ref{sec:interaction}), the properties of OB binaries are often considered as initial conditions and serve as input for other fields, such as population synthesis computations \citep{Eldridge+2017} or GW progenitor studies \citep{Belczynski+2016}. Multiplicity properties studied in the field as well as in cluster environments are described below.

Large magnitude-limited surveys of galactic OB stars in the field 
both in the Northern and Southern hemisphere such as IACOB or OWN showed that the observed binary fraction of O-type stars is around 55-65\% \citep{Abt1990, Sota2014, Barba2017, Britavskiy2019}. For B-stars in the field, the observed binary fraction was reported to decrease down to 45-20 \%, for early to late B-type stars, respectively \citep{Chini2012}. 

\begin{marginnote}[]
\entry{Binary fraction $f_\mathrm{bin}$}{The binary fraction is defined as $f\mathrm{bin} = N_B/N$ with being $N_B$ the number of objects with at least one companion and $N$ the number of objects. This can either be an observed or an intrinsic fraction.}
\end{marginnote}

Spectroscopic surveys of OB stars in young star clusters in the Milky Way found similar observed binary fractions than the ones reported in the field \citep{Sana2012, Kobulnicky2014, Banyard2022}. While they target only a small peculiar region, given the similar distance to all stars, they can be considered volume-limited samples. The VLT-FLAMES Tarantula Survey (VFTS), targeting the 30 Doradus star-forming complex in the LMC, \citep{Evans2011, Sana2013, Dunstall2015} measured lower observed binary fractions of $25\pm2\%$ to $35\pm3$\% for early-B and O-type stars. This is similar to the observed binary fraction measured in the young cluster NGC 346 in the SMC \citep{Dufton2019}. Using low-resolution spectroscopy, the observed binary fraction in NGC 330, a slightly older cluster in the SMC, was found to be even lower with only $\sim13\pm2$\% \citep{Bodensteiner2020_NGC}. 

In principle, the observed binary fractions give a lower limit on the intrinsic binary fractions. However, observed fractions are not directly comparable as different observing campaigns use different instruments and technical setups and their sensitivity to detect RV shifts thus varies. Moreover, there is a selection bias towards binary systems with similarly bright components rather than single stars at the faint end of magnitude-limited samples \citep{Vanbeveren1980}. To obtain the true, intrinsic binary fraction, the observed binary fractions have to be bias-corrected using the probability of detecting binary systems with a given observing campaign. This not only requires an assumption on the underlying orbital parameter distributions, but also detailed knowledge of potential biases of the observing campaign \citep[see e.g.,][]{Sana2012}. Comparing the intrinsic binary fractions (where available) demonstrates that a vast majority of massive stars are members of close binary systems. Despite large error bars, there seems to be a trend between binary fraction and stellar mass. Whether this is indeed related to the mass, or potentially also to the metalliticy or the age of the cluster 
remains to be constrained by further work.

Most surveys mentioned in Table\,\ref{tab:bin_stats} consider only the so-called SB1 bias (the chance to detect RV shifts of a given binary system larger than a given detection threshold, usually chosen to be 20\,km\,${\rm s}^{-1}$). An additional bias, the so-called SB2 bias, arises from unidentified SB2 systems that appear as rapidly rotating single stars if their RV shifts are too small for their spectral lines to effectively deblend \citep[see][]{Bodensteiner2021}. This effect becomes more important for more rapidly rotating stars, for smaller RV variations, as well as for lower spectral-resolution data. Taking this bias into account will further increase some of the intrinsic binary fractions reported in Table\,\ref{tab:bin_stats}.

While some of the surveys mentioned in Table\,\ref{tab:bin_stats} are based on few epochs only (enough to measure RV variations and detect binaries), others had enough observations (i.e., $\gtrsim$ 20 epochs) to fit binary orbits and constrain orbital parameters for SB1s and SB2s. Furthermore, additional observing campaigns were designed to follow-up previously detected binary systems, for example the MONOS survey for Galactic O-star binaries \citep{MaizAppellaniz2019, TriguerosPaez2021}, or the TMBM \citep{Almeida2017} and BBC programs \citep{Villasenor2021a} that followed O- and B-type binaries from the VFTS, respectively. 

The exponent of the period distribution, which is defined as $f(\log\mathrm{P[days]})\sim (\log\mathrm{P})^{\pi}$, is reported to vary between $\pi=-0.45 \pm 0.39$ for the Milky Way O stars \citep{Sana2012} and $\pi \sim -0.2$ for the LMC O stars \citep{Almeida2017}, which is close to flat on $\log P$ \citep[see e.g.,][]{Banyard2022}. The eccentricity distribution follows $f(e)\sim e^{\eta}$ with $\eta=-0.4\pm0.2$ \citep{Sana2012}. Covering a large range of mass ratios, \citet{Shenar2022} reported that the mass-ratio distribution of VFTS binaries is also consistent with a flat distribution, that is $f(q)\sim q^{\kappa}$ with $\kappa=-0.2\pm0.2$. Comparing the observed orbital parameter distributions derived from detected binaries in different spectroscopic works implies them to be universal across metallicity and the considered stellar mass range \citep[for a compilation, see figures 8 and 10 in][]{Banyard2022}. The orbital properties of close, eclipsing binaries in the Milky Way, LMC and SMC further corroborate that there are no statistically significant trends with metallicity \citep{Moe2013}.

Observations have unambiguously shown that a large majority of O- and B-type MS stars are in binary or higher-order multiple systems. While the binary fraction seems to increase with stellar mass also outside the here considered mass range \citep[for an overview, see e.g.,][]{Moe2017}, the uncertainties are still large. The orbital parameter distributions seem universal across masses and metallicity, but more observations are required to further test this. The measured binary properties in young clusters and environments are thought to reflect the initial conditions, but their link to the outcome of star formation is not well understood \citep{Duchene2013}. Early processes such as inward migration or binary hardening seem to also play a role \citep[e.g.,][]{RamirezTannus2021, Bordier2022}. The picture is further complicated by occurring interactions and post-interaction products polluting older populations and field stars \citep[see e.g.,][]{Wang+2020}.

\begin{table}
\caption{Spectroscopic surveys of OB stars investigating binarity. The columns give the number of observed stars n$_*$, the age of the population for clusters and associations, the observed ($f_\mathrm{bin}^\mathrm{obs}$) and bias-corrected ($f_\mathrm{bin}^\mathrm{intr}$) spectroscopic binary fraction, the period P$_{max}$ up to which the observational biases are corrected for, the name of the survey, the environment (env.) and relevant references (Ref.). They are sorted by host galaxy.}\label{tab:bin_stats}
\begin{center}
\begin{tabular}{l l l l l l l l l l} \hline
n$_\mathrm{*}$ & age & SpT & $f_\mathrm{bin}^\mathrm{obs}$ & $f_\mathrm{bin}^\mathrm{intr}$ & P$_\mathrm{max}$ & survey & Env. & Ref. \\ 
& [Myr] & & [\%] & [\%] & [days] & & & \\ \hline 
\multicolumn{9}{c}{Milky Way} \\ \hline
 205 & ? & O\&WN & 55 $\pm$ 3 & -- & & OWN & Southern Hem. & (1) \\
194 & ? & O & 65 $\pm$ 3$^\dag$ & -- & & GOSSS & mixed & (2) \\
319 & ? & O & $43 \pm 3^{+}$ & -- & & IACOB & Northern Hem. & (3)  \\
243 & ? & O & 68 $\pm$ 3 & -- & & BESOS & mixed & (4) \\
226 & ? & early-B & 46 $\pm$ 3 & -- & & -"- & mixed & (4) \\
353 & ? & late-B & 19 $\pm$ 2 & -- & & -"- & mixed & (4) \\ 
 74 & ? & B2-B5 & 35 $\pm$ 4 & -- & & & mixed & (5) \\
\hline
 71 & 1-4 & O3-O9.7 & 56 $\pm$ 6 & 69 $\pm$ 9 & 3000 & & clusters$^\dag$ & (6) \\
 80 & 2-7 & B0-B9 & 33 $\pm$ 5 & 52 $\pm$ 8$^*$ & 3000 & NGC 6231 & cluster & (7) \\ 
128 & 3-4 & OB & 51 $\pm$ 7 & $\sim$55 & 5000 & Cyg OB2 & association & (8) \\
 81 & 10-15 & B & 39 $\pm$ 9 & -- & 100 & Sco OB2 & association & (9) \\ \hline
\multicolumn{9}{c}{LMC} \\ \hline
360 & $5-15$ & O &35 $\pm$ 3 & 51 $\pm$ 4 & 3000 & 30 Dor & SF complex & (10)\\
408 & $5-15$ & early-B & 25 $\pm$ 2 & 58 $\pm$ 11 & 3000 & 30 Dor & SF complex & (11) \\ \hline
\multicolumn{9}{c}{SMC} \\ \hline
47 & $\sim$2 & O & 26 $\pm$ 6 & -- & & NGC 346 & cluster & (12) \\
288 & $\sim$2 & B & 18 $\pm$ 2 & -- & & -"- & cluster & (12)\\
284 & 35-40 & B & 13 $\pm$ 2 & 34$^{+8}_{-7}$ $^*$ & 3000 & NGC 330 & cluster & (13)  \\
\hline
\end{tabular}
\end{center}
\begin{tabnote}
$^\dag$ The observed clusters are IC\,1805, IC\,1848, NGC\,6231, NGC\,6611, Tr\,16 and IC\,2944. \newline
$^{+}$ We here combine the fractions for slow and rapid rotators reported by \citet{Britavskiy2019}.\newline
$^\ddag$ The fraction rises to 91 $\pm$ 2 \% when including possible detections.\newline
$^*$ Studies taking into account both SB1 and SB2 bias for the intrinsic binary fraction.\newline
References: (1) \citet{Barba2017}, (2) \citet{Sota2014}, (3) \citet{Britavskiy2023}, (4) \citet{Chini2012}, (5) \citet{Abt1990}, (6) \citet{Sana2012}, (7) \citet{Banyard2022}, (8) \citet{Kobulnicky2014}, (9) \citet{Levato1987}, (10) \citet{Sana2013}, (11) \citet{Dunstall2015}, (12) \citet{Dufton2019}, (13) \citet{Bodensteiner2021}
\end{tabnote}
 \end{table}

\subsection{OBe stars}
OBe stars appear on average cooler and redder than their OB counterparts because of their rapid rotation and the IR excess from the disk \citep[see][for a recent review]{Rivinius2013}. While the formation mechanism of the OBe disk remains debated, different channels explaining the rapid rotation have been proposed. Those include their interpretation as single stars that approach critical rotation towards the end of the MS \citep[e.g.,][]{Granada2013, Hastings2020}, or as mass gainers in previous binary interactions \citep[e.g.,][]{Pols1991, Hastings2021}. A potential way to distinguish those is to constrain OBe multiplicity properties: according to the single-star channel, OBe stars should have similar binary properties as OB stars. According to the binary channel, there should be a lack of MS companions and a significant number of stripped companions or compact objects.
\begin{marginnote}[]
\entry{Classical OBe}{Classical OBe stars are rapidly rotating, non-radially pulsating late-O and B-type stars whose spectra show, or have previously shown, emission lines that are formed in a gaseous circumstellar decretion disk.}
\end{marginnote}

Unfortunately, multi-epoch spectroscopic surveys constraining close binary properties of large samples of classical OBe stars are lacking, in particular in comparison to works on OB stars (see Section\,\ref{subsec:binstats_OB}). Additionally, OBe star spectra are complex and exhibit variability both in their disk emission and through stellar pulsations \citep[e.g.,][]{Barnsley2013, Baade2016, Labadie2022}. This complicates binary detections through spectroscopic and photometric techniques. 

\citet{Abt1978} studied a small sample of B and Be stars, reporting similar binary properties, however based on low-quality data. 
High-angular resolution imaging and speckle interferometry studies reported a similar result, but their observations are not sensitive to short periods relevant for binary interactions \citep{Oudmaijer2010, Horch2020}. \citet{Klement2020} interpreted a detected turn-down in the spectral energy distributions (SEDs) of OBe stars as the presence of a companion truncating the disk. The nature of these companions remains unconstrained, however. Using multi-epoch low spectral resolution observations of the young SMC cluster NGC 330, \citet{Bodensteiner2021} reported a significant lower binary fraction of Be stars compared to B-type stars in the same cluster.

Concerning Be star companions, based on a literature study, \citet{Bodensteiner2020_Be} pointed out a lack of close MS companions to massive Be stars\footnote{A possible exception might be $\delta$\,Sco, which was reported to be in a long-period, highly eccentric binary system with a MS companion. The system seems to be more complicated, however, as it shows indication of being a runaway triple system with another, yet undetected companion \citep{Miroshnichenko2001, Tango2009, Meilland2011}.} despite the fact that those companions are the easiest to detect and the most common companions to OB stars \citep{Shenar2022}. In contrast, the sample of Be binaries with bloated stripped stars and subdwarf OB (sdOB) companions is continuously increasing \citep[e.g.,][see Section\,\ref{subsec:stripped}]{WangL2021, ElBadry2021_HR6819}. A well-studied class of Be binaries are Be X-ray binaries (BeXRBs) that are detected based on their X-ray emission \citep[e.g., ][and see Section\,\ref{sec:nss}]{Raguzova2005, Reig2011, Coe2015}. 

Overall, the binary properties of OBe stars remain poorly constrained and the few measurements of orbital parameters are not enough to construct representative parameter distributions. The overall binary fraction of OBe stars seems to be lower than for OB stars while the companions are predominantly stripped stars or compact objects. This is in line with expectations of the binary channels of Be star formation, but further observations are required to consolidate or falsify this. On the lower-mass end (for late-type Be stars), the single-star channel might play a more important role \citep[e.g.,][]{Kervella2008, Klement2021}.

\subsection{Red supergiants}
While the number of red supergiants (RSGs) in the Milky is low, they also show a high degree of photospheric variability \citep[e.g.,][]{Kiss2006}, hampering the detection of RV variability and the signature of companions. Given their large radii of up to 1500\,$\mathrm{R_\odot}$, short-period binaries will not survive until the RSG phase and interact before, implying that there should be a lack of RSGs in close binaries, in contrast to their OB progenitors, and that several RSGs could actually be the rejuvenated products of previous interactions \citep[dubbed red stragglers,][]{Britavskiy2019}. 
\begin{marginnote}[]
\entry{Red supergiants}{Red supergiants are the evolved products of massive star evolution; for many stars they mark the final stage of evolution before they end their lives with a SN explosion}
\end{marginnote}

Given the rarity of RSGs in our Galaxy, multiplicity studies mainly targeted RSGs in our neighboring galaxies.
Compiling archival RV measurements of almost 1000 candidate RSGs in the LMC and SMC, \citet{Dorda2021} find a minimum binary fraction of $15\pm3\%$. 
RSGs with OB companions can be detected based on signatures in the blue part of the spectrum \citep{Neugent2018}. Using this method and spectroscopic follow-up observations, \citet{Neugent2019} measured a bias-corrected binary fraction of $20\pm7\%$ for RSGs in the LMC.  
\citet{Patrick2019} investigated multi-epoch spectroscopy of 17 candidate RSGs in the 30 Dor region in the LMC observed in the context of the VFTS survey. Given that the detection probability for systems with periods up to 10\,000 days and mass ratios $>0.3$ is estimated to be almost 90\%, the observed and intrinsic binary fractions are similar, namely $\sim30\%$. \citet{Patrick2020} further investigated the binary properties of 15 RSGs in the 35-40\,Myr old SMC clusters NGC~330 with multi-epoch spectroscopy. They report a bias-corrected binary fraction of $30\pm10\%$ for orbital periods between $2.3< \log\mathrm{P[days]}< 4.3$ and mass ratios $>0.1$.

Based on UV photometry, \citet{Patrick2022} measured an intrinsic binary fraction of $18.8\pm1.5\%$ for $>500$ RSGs in the SMC for mass ratios $>0.3$ and long periods above 1000 days. They find indications for a flat mass-ratio distribution, and a lack of high mass-ratio ($q>0.5$) systems above 15\,$\mathrm{M_\odot}$ (implying an absence of companions with similar masses to that of the RSG), which the authors interpreted as the most massive RSGs being merger products. \citet{Neugent2021} extended the investigation to the Local Group galaxies M31 and M33. They reported a constant binary fraction of $33.5^{+8.6}_{-5.0}\%$ for M31, while the binary fraction in M33 drops from $41.2^{+12.0}_{-7.3}\%$ to $15.9^{+12.4}_{-1.9}\%$, for inner and outer regions, respectively, which is attributed to a metallicity effect.

Overall, the intrinsic binary fractions of RSGs measured in the SMC and LMC using different detection techniques are in agreement with each other and amount to $\sim30\%$. The role of metallicity, the nature of the companions of those RSGs as well as the orbital parameter distributions remain to be further constrained.

\subsection{Luminous Blue Variables}
While LBVs were for a long time considered to be the transitory phase between single-star BSGs and WR stars characterized by strong, eruptive mass loss \citep[e.g.,][]{Conti1975, Lamers2002}, the link of LBV as direct progenitors of type-II SNe induced a shift in their interpretation, not as transitory phase but as potential end-point of stellar evolution \citep{Kiewe2012, Groh2013}. They were further proposed to not be a phase in single-star evolution, but to be binary interaction products \citep[see see e.g.,][and Section\,\ref{sec:post}]{Gallagher1989, Justham2014, Smith2015}.
Constraining the binary properties of LBVs would thus also allow to put constraints on their evolutionary history. 
\begin{marginnote}[]
\entry{Luminous Blue Variable}{Luminous blue variables (LBVs) are in a brief phase in the evolution of evolved massive stars. As the name implies, they populate the upper region of the HRD and are hot, can be above the HD limit, and are highly variable.}
\end{marginnote}

Given the rarity of LBVs and their strong (spectroscopic and photometric) variability \citep[see e.g.,][]{Humphreys1994}., binary detections are difficult. Targeting individual objects with different methods, several binary LBVs have been found so far \citep[see e.g.,][]{Martayan2012, Boffin2016}. Only few systematic studies of the binary properties of LBVs exist. Based on an X-ray survey using XMM-Newton, \citet{Naze2012} report X-ray properties for LBVs consistent with a binary a fraction between 26-69\% (detecting signatures of wind-wind collisions and potentially the O-type companions themselves). Targeting seven LBVs with high-angular resolution imaging, \citet{Martayan2016} find a binary fraction of $\sim 30\%$. \citet{Mahy2022} combined spectroscopy and interferometry and reported a bias-corrected binary fraction for galactic LBVs of $62^{+38}_{-24}\%$ for periods below 1000 days and mass ratios between 0.1 and 1.0, higher than previously expected. The detected companions of LBVs are either OB MS stars or RSGs, and given the large radii of LBVs the periods of those systems are large. The full orbital parameter distributions such as the period distribution remain largely unconstrained. 

\subsection{Wolf-Rayet stars}
Different types of stars fall under the spectroscopic definition of Wolf-Rayet (WR) star, which is based on strong, broad emission lines in the spectrum. On the one hand there are classical WR (cWR) stars, sub-divided in nitrogen- (WN), carbon- (WC) or oxygen-rich (WO) depending on their composition. Most WR stars fall in the cWR category and are thought to be post-H-burning objects in an evolved stage of massive-star evolution after the RSG and potentially LBV phase. On the other hand, most of the rare WNh stars (nitrogen-rich WR stars with hydrogen lines) are thought to be even more massive stars ($\gtrsim100\,\mathrm{M_\odot}$ at galactic and LMC metallicity) on the MS with strong stellar winds \citep[see e.g.,][for a recent review]{Crowther2007}. For those, \citet{Conti1975} proposed an evolutionary sequence of O/WNh $\rightarrow$ LBV $\rightarrow$ WN $\rightarrow$ WC ($\rightarrow$ WO). If this evolutionary connection indeed exists, their binary properties should also align in this sequence (i.e., remain the same or decrease in later stages).
\begin{marginnote}[]
\entry{Wolf-Rayet stars}{WR stars are classified based on their spectra that are dominated by strong, broad emission lines, and comprise a number of objects with a different origin.}
\end{marginnote}

In general, and using a multitude of techniques, similar observed binary fractions of $\sim$30-40\% were reported for the Milky Way \citep[e.g.,][]{vanderHucht2001, Dsilva2020, Dsilva2022, Dsilva2023}, lower-metallicity environments \citep[such as the LMC and SMC, e.g., ][and references therein]{Foellmi2003a, Foellmi2003b, Schnurr2008} and higher-metallicity host galaxies \citep[like the Triangulum or the Andromeda Galaxy,][]{Neugent2014}. High-resolution spectroscopy of Galactic WN and WC stars showed that the bias-corrected binary fraction of WN stars is lower ($0.52^{+0.14}_{-0.12}\%$) than for WC stars ($0.96^{+0.04}_{-0.22}\%$). Also their period distributions differ, with a large number of short-period WN binaries \citep[][and references therein]{Dsilva2023}. This implies that while the evolutionary connection described above might hold for long-period systems, the short-period WN systems remain unexplained. Additionally, binary properties of WNh stars remain largely unconstrained.

\subsection{Blue and yellow supergiants}
The class of blue supergiants (BSGs) comprises several types of objects: O supergiants are massive stars towards the end of their MS evolution. B supergiants are interpreted as a transitory post-MS phase, either in a brief traverse over the HRD or during a blue loop \citep[][]{Maeder2001, Vink2022}. Long-lived BSGs were suggested to be the product of stellar mergers. One particular case is that of SN1987A, which occurred in the LMC. Pre-explosion images indicated its progenitor was a BSG, which can be naturally explained as a merger product \citep{Podsi1990, MenonHeger2017}.
Yellow supergiants (YSGs) are a heterogeneous group of stars crossing the HRD \citep{Maeder2000}.

The binary properties of BSGs and YSGs are not well constrained. In the context of the VFTS, \citet{Dunstall2015} reported an observed binary fraction ob B-type supergiants of $23\pm6\%$, in agreement with the one observed for MS B-type stars targeted in the same survey. Observing OB giants to supergiants in the Galactic clusters Westerlund\,1 with VLT-FLAMES, \citet{Ritchie2022} found an observed binary fraction of $\gtrsim$40\%, with an intrinsic binary fraction that might be significantly larger. 

\section{BINARY INTERACTION PROCESSES}\label{sec:interaction}
The interactions that will happen (or have already happened) for the binary systems discussed in the previous section are mainly dictated by the radial evolution of both components. As a star evolves, it can eventually fill its Roche lobe, producing outflows through the first Lagrangian point L1 towards its companion. In binary evolution models, Roche lobes are approximated by a volume-equivalent radius, normally computed through the fit made by \cite{Eggleton1983} in terms of the mass ratio\footnote{Taking $m_d$ and $m_a$ as the mass of the donor and accretor star respectively, mass ratios in literature are often defined as either $q\equiv m_d/m_a$ or its inverse. For clarity, it is always best to specify the definition of $q$ used.} ($q\equiv m_1/m_2$):
\begin{eqnarray}
R_\mathrm{RL,1}(q) = f(q)a,\quad f(q) =\frac{0.49 q^{2/3}}{0.6q^{2/3}+\ln(1+q^{1/3})},
\end{eqnarray}
where $R_\mathrm{RL,1}$ is the Roche lobe radius of the star with mass $m_1$, and $a$ is the orbital separation. Even if a star does not fill its Roche lobe, interactions can still happen through tidal effects, wind-orbit coupling and irradiation. On the opposite extreme, binary stars can extend well beyond the L1 point, either in stable contact systems or dynamically evolving common envelope (CE) configurations.

Roche lobe overflow (RLOF) is the most transformative form of interaction in binaries. Following the work of \citet{KippenhahnWeigert1967}, phases of mass transfer are nominated as ``Cases'' (capitalized for historical reasons). Case A mass transfer refers to RLOF episodes during the MS of the donor star, while Case B mass transfer refers to RLOF before core-helium depletion. Case B is often split into early and late Case B, indicating interaction before or after the development of a convective envelope. Case C mass transfer corresponds to late interactions post core-helium depletion. In this review we make use of these definitions but note that they can be inconsistent in the literature, with Case B and Case C representing instead mass transfer before and after the ignition of helium, respectively \citep{Podsiadlowski1992}. Repeated interactions are marked with multiple letters, such that Case AB refers to a mass transfer event after the MS, which was preceded by Case A mass transfer.

\begin{figure}
    \centering
    \includegraphics[width=\columnwidth]{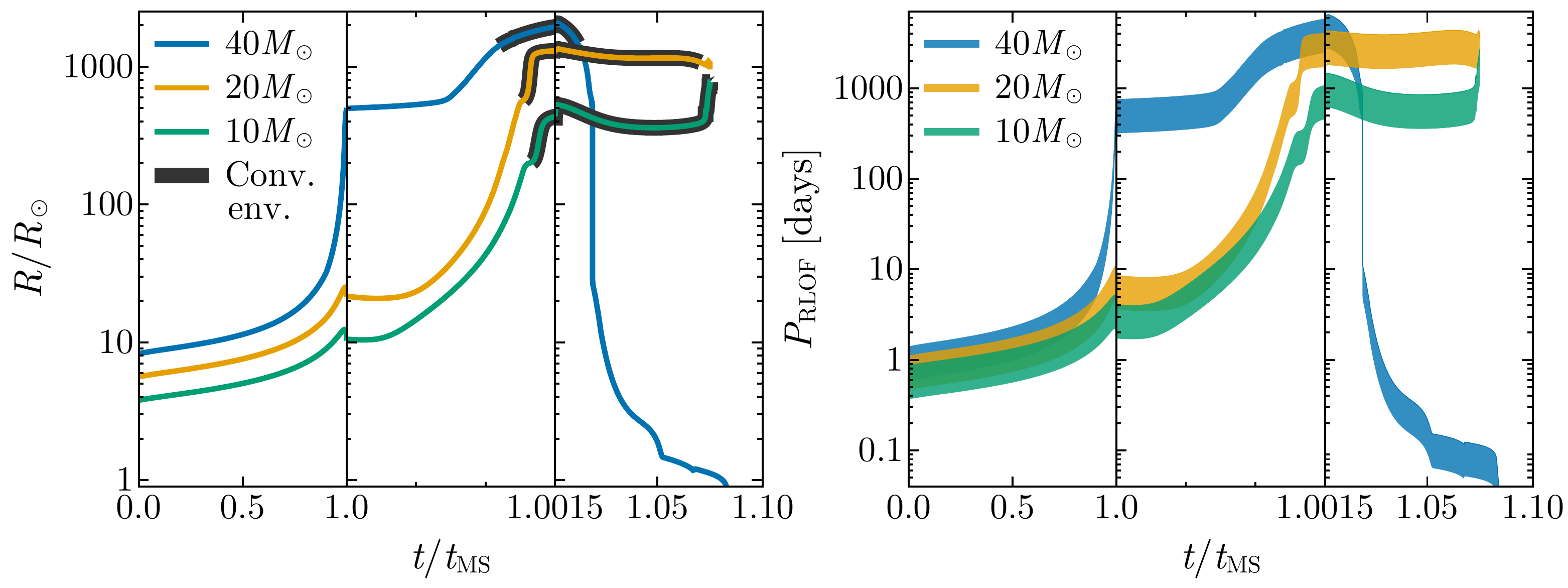}
    \caption{\textit{(left)} Radial evolution of massive stars at solar metallicity with different masses (see legend) as a function of time in units of the MS lifetime. The thick black line indicates the points at which the star develops a convective envelope containing more than 10\% of the total mass. \textit{(right)} Range of orbital periods (considering an arbitrary mass ratio) for which these stars would undergo Roche lobe overflow.}
    \label{fig:star_interaction}
\end{figure}

The range in separations and orbital periods at which RLOF is expected is illustrated in Figure \ref{fig:star_interaction}. As massive stars can expand by over two orders of magnitude, RLOF can occur at orbital separations ranging from tens of solar radii up to about ten AU (for mass ratios of order unity). For a given stellar radius $R_1$, the period at which RLOF would occur is given by
\begin{eqnarray}
    P_\mathrm{RLOF} = R_1^{3/2}\left(\frac{4\pi^2}{Gm_1}\right)^{1/2}g(q), \quad g(q)\equiv (f(q))^{-3/2}\left(\frac{q}{1+q}\right)^{1/2}.
\end{eqnarray}
The function $g(q)$ is bound between $3.3$ and $1.35$, resulting in a finite range of periods for which a star would undergo RLOF, independent of the mass ratio. As shown in the left panel of Figure \ref{fig:star_interaction}, solar-metallicity massive stars are expected to undergo RLOF at the zero-age MS for periods $\sim 1$ day, while at their maximum expansion they can undergo RLOF at periods of $\sim 10$ years. As such, interacting binary systems operate over a large range of spatial and temporal timescales.

Orbital evolution can be computed in terms of the orbital angular momentum,
\begin{eqnarray}
    J=m_1m_2\sqrt{\frac{Ga(1-e^2)}{m_1+m_2}}.
\end{eqnarray}
It is common to assume near-RLOF systems to be efficiently circularized (eg. \citealt{VerbuntPhinney1995}), in which case the time derivative of the orbital angular momentum is given by
\begin{eqnarray}
\frac{\dot{J}}{J}=\frac{\dot{m}_1}{m_1} + \frac{\dot{m}_1}{m_2} - \frac{1}{2}\frac{\dot{m}_1+\dot{m}_2}{m_1+m_2} + \frac{1}{2}\frac{\dot{a}}{a}.\label{equ:J}
\end{eqnarray}
Having a model for how the masses of a binary system evolve with time and how angular momentum is removed or added to the orbit allows for the integration of this equation to determine the evolution of the orbital separation. For some simple models the evolution of the separation as a function of the component masses can be determined analytically, serving as a useful probe to determine the future outcome of a binary observed pre-RLOF, or to understand the potential progenitors of a post-interaction system (see, eg., \citealt{Soberman+1997,TaurisVandenHeuvel2023}). The detailed dynamics of binary outflows and the angular momentum they remove remains an important uncertainty in evolutionary calculations (eg. \citealt{BrookshawTavani1993,MacleodLoeb2020,Willcox+2023}).

\subsection{Modeling tools for binary evolution}
Following the evolution of a binary system, including the properties of both its components, requires the use of computer simulations. Three different types of codes are used for this purpose. (Magneto)-hydrodynamical simulations can be used to probe short dynamical timescales, useful to model processes such as mergers and CE evolution (eg. \citealt{Lombardi+1995, TaamSandquist2000, Schneider+2019,Lau+2022}). Due to their computational cost 3D simulations can only explore a limited set of initial conditions, and cannot resolve nuclear- and thermal-timescale mass-transfer processes. To model longer evolutionary timescales, 1D stellar evolution codes are used, which at their core carry a significant resemblance to the models computed over half a century ago by \citet{KippenhahnWeigert1967}. 1D stellar evolution codes also serve as the source for initial conditions of 3D simulations. The third type of simulations are referred to as ``rapid'' codes, and are based on semi-analytical models based on pre-computed 1D stellar models (eg. \citealt{Hurley+2002}) which allows for sub-second calculation of full binary evolution models.

\begin{marginnote}[]
\entry{Detailed binary code}{1D stellar evolution code that solves the differential equations of stellar structure and evolution for a binary system. Has a runtime of order a cpu hour.}
\end{marginnote}
\begin{marginnote}[]
\entry{Rapid binary code}{Code based on fits to single star evolution models and semi-analytical approximations to binary interactions. Has runtimes smaller than a cpu second.}
\end{marginnote}

In contrast to rapid evolutionary codes, 1D simulations are referred to as ``detailed''. The majority of detailed binary models currently computed are done with either the \texttt{MESA} code \citep{Paxton+2011,Paxton+2015} or variations of the \texttt{STARS} code \citep{Eggleton1971, Eldridge+2008}. These codes are derived from the Henyey method \citep{Henyey+1959} to solve the equations of stellar structure in 1D, computing the evolution of two stars that are coupled through tidal interaction (see Section \ref{sec:tidal}) and mass transfer (see Section \ref{sec:mt}). Detailed models can accurately follow phases of evolution on the nuclear and thermal timescales of its components, including phases of contact (see Section \ref{sec:contact}). Using 1D codes to model dynamical phases of evolution requires approximations that ignore (or parameterize) the uncertainties associated with 3D binary interactions (see Section \ref{sec:unstable}).

Compared to detailed evolutionary codes, a much broader set of tools are available to perform rapid calculations. The majority of these are based on the analytical fits of \citet{Hurley+2000} to the single star evolution models of \citet{Pols+1998}, coupled with the semi-analytical approximations to binary evolution of \citet{Hurley+2002}. These include \texttt{Startrack} \citep{Belczynski+2002}, \texttt{binary\_c} \citep{Izzard+2004}, \texttt{MOBSE} \citep{GiacobboMapelli2018}, \texttt{COSMIC} \citep{Breivik+2020} and \texttt{COMPAS} \citep{Riley+2022}. Even though these codes include multiple free parameters to adjust the physics of binary evolution, they are limited to the physical assumptions of the stellar models of \citet{Pols+1998}. Some examples that break from this reliance on the fits of \citet{Hurley+2002} are \texttt{COMBINE} \citep{Kruckow+2018}, \texttt{METISSE} \citep{Agrawal2020} and \texttt{SEVN} \citep{Iorio+2023}. Additionally, various codes precede the work of \citet{Hurley+2002}, including \texttt{SeBa} \citep{PortegiesZwartVerbunt1996} and the Brussels population synthesis code \citep{Vanbeveren+1998}. One main weakness of rapid evolutionary codes is that, since they are based on single-star evolutionary models, they cannot capture the response of thermal timescale mass transfer on either component, which plays a critical role in determining the stability of mass transfer.

Owing to their short runtimes, rapid codes have been the preferred tool to perform population synthesis calculations (see \citealt{Han+2020} for an overview), but detailed calculations are continuously taking a larger role in this area. Large grids of detailed models have been usually restricted to limited regions of the input parameters (eg. the grids of case A evolution of \citealt{NelsonEggleton2001}, \citealt{deMink+2007} and \citealt{Sen+2022}). Currently, the \texttt{BPASS} \citep{Eldridge+2017} and \texttt{POSYDON} \citep{Fragos+2023} codes (which use \texttt{STARS} and \texttt{MESA} as their backends, respectively) provide openly available sets of detailed calculations covering the full range of parameters relevant to interacting massive binaries. Other large-scale population synthesis calculations done with detailed models have been performed (eg. \citealt{Wang+2020}) but do not have an associated name. Although detailed population synthesis calculations are now feasible, further increasing the efficiency of calculations is critical for reproducibility and broad testing of theoretical uncertainties. One approach is the use of smart sampling of initial parameters rather than the use of a regular grid, coupled with interpolation \citep{Rocha2022}.

\subsection{Tidal Interaction}\label{sec:tidal}
Tidal torques in stars with radiative envelopes are attributed to the dynamical tide process (\citealt{Zahn1975}, see \citealt{Zahn2008} for a recent review). In this process, gravity modes are excited by the tidal potential near the interface between the convective core and the radiative envelope, with these waves dissipating close to the stellar surface. The work of \citet{Zahn1975} provides a straightforward method to compute the rate of tidal synchronization and circularisation, with the only non-trivial dependency being the computation of the structure constant $E_2$. Many simulations \citep{Hurley+2002,Paxton+2015} relied on an interpolation to the values of $E_2$ for zero-age MS models computed by \citep{Zahn1975}, but more modern calculations accounting for evolved stages are available \citep{Qin+2018}. Extensions to the \citet{Zahn1975} model have also shown the potential for resonant interactions when the tidal frequency matches a natural oscillation frequency of the star, potentially leading to tidal locking \citep{WitteSavonije1999}. Direct calculations of the tidal torques are now possible on a timescale that allows for their integration in evolutionary calculations (eg. \citealt{Sun+2023}). Computations of this type, relaxing some of the assumptions made by \citet{Zahn1975}, have shown that under some circumstances, tidal interaction arises from standing waves rather than traveling waves that are completely damped at the surface \citep{MaFuller2023}.

An associated growing field of study is the observation and theory of tidally excited oscillations (TEOs) in eccentric binaries (eg. \citealt{Fuller+2017} and references within). The discovery of the prototypical system KOI-54 \citep{Welsh+2011} was enabled by the short cadence and high precision observations of the \textit{Kepler} telescope. Owing to their characteristic lightcurves, such systems are often referred to as heartbeat stars. The two components of KOI-54 are only about twice the mass of the Sun, but despite not being massive they allow to probe tidal processes in stars with radiative envelopes. Further observations have pushed the observed heartbeat systems to the massive star regime, with TESS observations showing a system with a total mass of $\sim 150\;\mathrm{M_\odot}$ \citep{Kolaczek-Szymanski+2021}.

One aspect that has received less attention is the effect of tidal deformation. Although tidally deformed stellar surfaces following the Roche potential are used to model observations (eg. \citealt{PrsaZwitter2005, Abdul-Masih+2020}), the impact of deformation on interior structure is seldom included in evolutionary models. By extending methods used to model centrifugal deformation in 1D stellar models, \citet{Fabry+2022} has incorporated tidal deformation in binary evolution models, which can be applied to detached, semi-detached and contact binary systems. Another approach by \citet{FellayDupret2023} accounts for the full non-spherical mass distribution of each component in order to construct static structure models. Extended studies describing how tidal deformation modifies binary evolution are not available at the moment.

\subsection{Stable mass transfer}\label{sec:mt}
When a star in a binary fills its Roche lobe, mass transfer will ensue and modify both the orbit and the structure of the donor. Whether or not the donor can remain in hydrostatic equilibrium in these conditions is normally described in terms of the mass radius exponents,
\begin{eqnarray}
\zeta_\mathrm{ad}\equiv\frac{d \log R}{d\log M}_\mathrm{ad}, \quad \zeta_\mathrm{RL}\equiv \frac{d \log d R_\mathrm{RL}}{\log M},
\end{eqnarray}
which describe the adiabatic response of the donor radius and the change in the Roche lobe radius as mass is transferred (eg. \citealt{Soberman+1997}). If $\zeta_\mathrm{ad}>\zeta_\mathrm{RL}$, the donor can adjust to mass transfer while remaining in hydrostatic equilibrium, producing either nuclear or thermal timescale events. In particular, donor stars with a significant fraction of their mass in a convective envelope are expected to expand in response to mass loss \citep{HjellmingWebbink1987}, favoring instability for late Case B and Case C mass transfer (see Figure \ref{fig:star_interaction}). In detailed evolutionary calculations, this criteria is less useful, as $\zeta_\mathrm{ad}$ cannot directly be inferred from the structure of the star. In certain cases the pressure scale height at the photosphere can be an important fraction of the stellar radius, making the concept of a hard surface limited by its Roche lobe inapplicable. Such conditions require a different method to evaluate instability \citep{Temmink+2023}. Rapid evolutionary codes rely on prescribed stability criteria, leading, for instance, to significant uncertainties in the predicted rates and formation processes of merging binary black holes \citep{Olejak+2021, Gallegos-Garcia+2021}.

\begin{figure}
    \centering
    \includegraphics[width=\columnwidth]{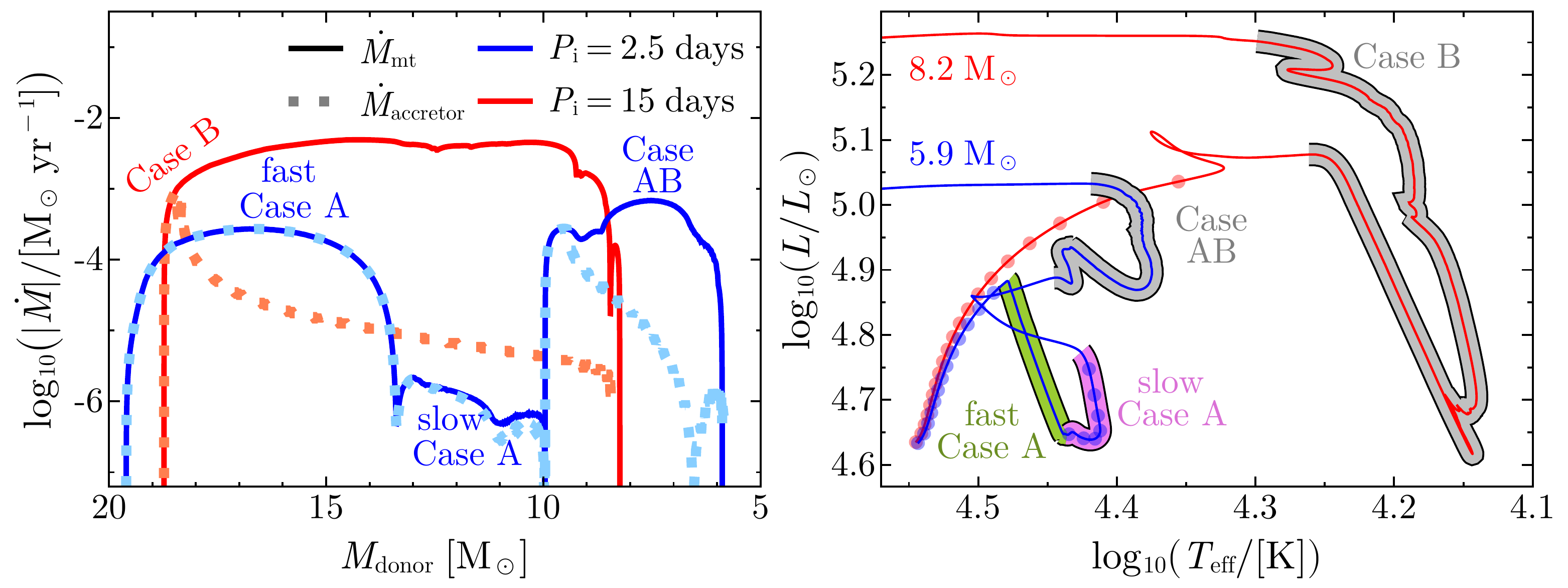}
    \caption{(left) Mass transfer rate $\dot{M}_\mathrm{mt}$ and rate of change of mass of the accreting star $\dot{M}_2$ as a function of the donor mass for a case A system ($20\,\mathrm{M_\odot} + 16\mathrm{M_\odot}$ with an initial period of 2.5 days) and an early Case B system (same masses but with an orbital period of 15 days). Simulations include rotation and mass accretion is limited after the accreting star reaches critical rotation. (right) Evolution of the donor star in the HR diagram for the same two binaries. Different mass transfer phases are indicated with thick contours, and dots are placed in time intervals equal to 5\% of the main-sequence lifetime of the donor star.}
    \label{fig:binary_rotation}
\end{figure}

Figure \ref{fig:binary_rotation} illustrates the typical evolution of two binaries undergoing stable Case A and early Case B mass transfer, closely resembling the evolution of the donor stars in \citet{KippenhahnWeigert1967}. Unless the evolution is modified by the secondary filling its own Roche lobe, Case A mass transfer is expected to be separated into a ``fast'' and a ``slow'' phase. As the more massive star in the Case A system initiates mass transfer, the orbit shrinks and leads to a thermal timescale mass transfer. After the mass ratio is inverted and the donor can thermally relax into the size of its Roche-lobe, mass transfer is driven by nuclear evolution. This evolutionary stage is referred to as Algol phase, and owing to the large ratio between thermal and nuclear timescales, a large majority of observable semi-detached systems are expected to be in this phase. After the MS, a final phase of Case AB mass transfer strips most of the remaining envelope and leaves a star mostly composed of helium. The Case B system instead interacts while its expansion is driven by hydrogen shell burning and results in a single fast phase of Case B mass transfer removing most of the hydrogen envelope.

There are various processes that can modify this classical picture of the evolution of massive binary stars. Rapid rotation is argued to produce chemically homogeneous evolution in massive stars \citep{Maeder1987}, in which case the MS evolution proceeds at almost constant radius. \citet{deMink+2009} argued that for massive binaries born near contact, tidal synchronization leads to rapid rotation and chemically homogeneous evolution of the primary, resulting in the first mass transfer phase being initiated by the initially less massive star (see also \citealt{Marchant+2017}). Regarding the Algol phase, \citet{Sen+2023} has shown that as more massive stars have a larger fraction of their mass in their convective cores, for sufficiently high donor masses ($\gtrsim 30 \mathrm{M_\odot}$) the helium-enriched core can be exposed before the mass ratio inverts. If that happens, the star can thermally relax to its Roche lobe size before mass ratio inversion, leading to a slow case A phase with a more massive donor than accretor (which \citealt{Sen+2023} refer to as an inverse Algol). Regarding Case B mass transfer, \citet{Klencki+2022}
has suggested that at metallicities smaller or equal to that of the LMC, the thermal mass transfer phase can be interrupted, and followed by a long-lived nuclear timescale phase with blue or yellow supergiant donors.

Mass accretion onto the secondary results in increased rotational velocities as well as contamination with CNO processed material from the donor star. These are considered key indicators of past binary interaction, but unfortunately are also degenerate with expectations of rotational mixing. As post-interaction secondaries can become overluminous and evolve to long orbital periods (or become single stars in case the primary undergoes a SN), apparently single stars can actually be predominantly post-interaction products \citep{deMink+2014}. Our theoretical understanding of CNO enrichment and spin-up through accretion has not evolved significantly in the past decade, and the reader is referred to sections 3.1 and 3.2 of \citet{Langer2012}.

One critical aspect that remains unsolved is the interplay between mass transfer efficiency (ie. how much of the mass transferred by the donor is accreted onto the companion) and accretion spin-up. As shown by \citet{Packet1981}, an accreting star needs only to increase its mass by a few percent before reaching critical rotation. Whether or not accretion can proceed from that stage is uncertain. \citet{Langer+2003} has pointed out that in the shortest period systems tidal interactions can prevent the accretor from reaching critical rotation, allowing for further accretion (see also \citealt{Sen+2023}). It has also been argued that angular momentum could be transported outwards from the accretion disk while still allowing for an inwards mass flow \citep{Paczynski1991,PophamNarayan1991}. Observational constraints are usually restricted to either the post-interaction or the Algol phase, while non-conservative phases are likely associated to thermal timescale mass transfer. Algol systems in the SMC appear to support a lowered efficiency with initial orbital period \citep{deMink+2007b} but other post-interaction systems favor high accretion efficiencies even at long orbital periods (eg. \citealt{Schootemeijer+2018, Bodensteiner2020_HR6819,Vinciguerra+2020}). A keystone system to understand accretion efficiency is the massive binary $\beta$ Lyrae (eg. \citealt{Mourard+2018}), which is currently undergoing a rapid mass-transfer phase and has been resolved with the CHARA interferometer.

Another limitation of current detailed binary models is that mass transfer rates are determined from prescriptions that use the 1D model structure of the donor star. A commonly used prescription is that of \citet{KolbRitter1990}, which treats separately the contributions from the extended atmosphere of the donor stars as well as from regions below the photosphere that are above the L1 equipotential. Lightening the assumptions of the \citet{KolbRitter1990} model \citep{Marchant+2021} or taking a different approach altogether to the computation of mass transfer rates \citep{CehulaPejcha2023} leads to qualitatively different evolution during fast mass transfer phases, with the potential to undergo overflow of the outer Lagrangian points.

\subsection{Contact Binaries}\label{sec:contact}
Contact binaries, where both components extend beyond the L1 equipotential, are precursors to stellar mergers and represent the most compact binary configurations possible. So long as material is contained within the equipotential of the second Lagrangian point L2, hydrostatic equilibrium is possible \citep{Kuiper1941}. Just as with rotating stars, hydrostatic equilibrium in radiative layers requires the radiative flux to be proportional to gravity (eg. \citealt{Fabry+2022}). However, the von Zeipel paradox \citep{vonZeipel1924} does not allow for both radiative and thermal equilibrium to hold simultaneously, such that large scale flows arise \citep{SmithSmith1981,TassoulTassoul1982} and the surface flux deviates from a simple proportionality to effective gravity (eg. \citealt{EspinosaLaraRieutord2012}). Contact binaries are expected to have similar temperatures at their surfaces, implying that the luminosity ratio between the two components is similar to the mass ratio ($L_1/L_2\simeq m_1/m_2$, \citealt{Lucy1968}). Owing to the steepness of the mass-luminosity relationship, this requires a significant redistribution of the luminosity beyond the L1 equipotential.

The internal structure of massive contact binaries, including tidal deformation and energy transport, is not well understood, with early work from the 70s not reaching a consensus (see \citealt{Shu+1980} and references within). Evolutionary models have characterized the conditions under which contact evolution happens, including the expansion of the accretor during thermal-timescale mass transfer as well the case where the more massive secondary in an Algol system expands due to nuclear evolution \citep{Eggleton1996, NelsonEggleton2001,WellsteinLanger+2001}. Modeling the evolution after that stage has remained uncertain, with models that either ignore the different physics of the contact stage, or use an ad-hoc mass transfer rate determined such that the surface of both stars remains in the same equipotential \citep{deMink+2007,Marchant+2016}. 
Population synthesis calculations of the LMC using this mass-transfer model indicate an overestimation of massive contact systems with mass ratios $\simeq 1$ when compared to observations \citep{Menon2021}, as most contact systems with mass ratios away from unity are predicted to evolve in a thermal timescale towards equalization of masses. Initial calculations by \citet{Fabry+2022,Fabry+2023} have shown that the inclusion of energy transport in evolutionary models of massive contact binaries can extend long-lived phases of evolution with mass-ratios away from unity.

As the components of a short period contact binary rotate rapidly, rotational mixing is expected to make them overluminous and rich in nitrogen at their surface (and potentially operate more efficiently than in single star evolution, \citealt{Hastings+2020b}). Observations of the contact binary VFTS 352 \citep{Almeida+2015}, containing a $\sim30\,\mathrm{M_\odot}+30\,\mathrm{M_\odot}$ binary with a period of 1.1 days, showed the system was indeed overluminous, providing potential support for rotational mixing and chemically homogeneous evolution. However, through detailed spectroscopic analysis accounting for the variable effective temperature across the surface, \citet{Abdul-Masih+2019,Abdul-Masih+2021} did not find an indication of enrichment with CNO processed material. Whether or not rotational mixing is active in massive contact stars remains an open question. The systems analyzed by \citet{Abdul-Masih+2019,Abdul-Masih+2021} show that both components share similar effective temperatures and luminosities, which is consistent with the scenario where energy is efficiencly transferred across the shared layers. \citet{Abdul-Masih+2022} also showed that for a selection of contact binaries with data spanning more than a decade, the evolution of their orbital periods could be constrained to operate on their nuclear timescale, independent of their mass ratio.

\subsection{Mergers and common envelope evolution}\label{sec:unstable}
As a contact binary grows beyond its L2 equipotential, outflows removing large amounts of angular momentum are expected to lead to a quick coalescence \citep{Pejcha+2016}. A stellar merger can naturally be confused with a single star, but current binaries could also be the result of a merger in a triple system as has been suggested for $\eta$ Car \citep{Hirai+2021}, HD 45166 (known as the quasi-Wolf Rayet star, \citealt{Shenar+2023}) and even higher multiplicity systems \citep{VignaGomez+2022}. It has been argued that the origin of magnetic stars is associated to amplification processes during a stellar merger \citep{Ferrario+2009}. Magneto-hydrodynamical merger simulations by \citet{Schneider+2019} have shown that indeed sufficiently strong fields are produced, and identified the amplification process to be the magneto-rotational-instability. However, even if sufficiently strong magnetic fields are produced, they could be short-lived. Strong magnetic fields introduce an additional timescale, associated to Alfven waves travelling through the star, and there are significant restrictions to the field geometry to be stable in this timescale \citep{BraithwaiteSpruit2004}.

A merger can also be the outcome of unstable mass transfer leading to CE evolution. Whether or not a binary system undergoing CE will survive (ejecting its shared envelope) or merge is a long-standing problem in binary evolution (see \citealt{Ivanova+2013} for a review). Being one of the main formation channels proposed to form binary neutron stars (eg. \citealt{Tauris+2017, Vigna-Gomez+2018}), and merging binary black holes (eg. \citealt{Belczynski+2016, Bavera+2021}), uncertainties in CE evolution lead to order-of-magnitude uncertainties on compact object merger rates \citep{MandelBroekgaarden2022}.

As individual CE simulations of massive stars are expensive (eg. \citealt{Lau+2022}), producing population predictions is reliant on simplified approximations. The most common approach is the use of the ``energy balance criterion'', where the change in orbital separation is computed in terms of the binding energy of the envelope of the star inititating the CE phase, and a free efficiency parameter \citep{Webbink1984}. Rapid population synthesis codes make use of fits to pre-computed binding energies as a function of the evolutionary stage of the star, but it has been recently pointed out that these could severely underestimate binding energies and overpredict the amount of stars surviving CE evolution \citep{Klencki+2021,Marchant+2021}. The computation of binding energies is also very uncertain, as it is not known a priori how much mass will be ejected before the CE. Recent results indicate that post-CE, the donor does not necessarily contract to become a hot stripped star, but rather undergoes a phase of stable mass transfer which completes the envelope stripping process \citep{Fragos+2019, Marchant+2021, HiraiMandel2022}.



\section{NON-DEGENERATE POST-INTERACTION PRODUCTS}\label{sec:post}

As described in Section\,\ref{sec:binarity}, a large number of stars in close binaries are predicted to interact at some point during their evolution. This not only drastically changes their evolution, but also implies that there should be a large number of post-interaction products \citep[e.g.,][]{deMink+2014, Schneider2015, Wang+2020}. The different evolutionary channels described in Section\,\ref{sec:interaction} can lead to a multitude of interaction products with different properties, depending on the type of interaction. While there are several characteristics proposed to identify interaction products \citep[see e.g., ][for a list of observational indications]{deMink+2014}, none of them are unambiguous, making their identification difficult. This implies that a large number of undetected interaction products might be interpreted as single- or non-interacting stars. As pointed out by \citet{deMink+2014}, the 'best' single stars are most likely stars in pre-interaction binaries, which can usually be detected by large RV variations (see Section\,\ref{subsec:binstats_OB}).

Including interaction products in our understanding of stellar populations is crucial. For example, binary interactions were proposed to be responsible for the split MSs observed in young star clusters \citep{Milone2018, Wang+2020, WangC2022}. Furthermore, stars that have undergone binary stripping during their evolution are predicted to significantly impact the integrated spectrum of stellar populations \citep{Gotberg2019} and change their ionizing budget \citep{Gotberg2020}. Stripped stars are further thought to produce systematically different SN yields \citep{Laplace2021}. 

\begin{figure}
    \centering
    \includegraphics[width=\columnwidth]{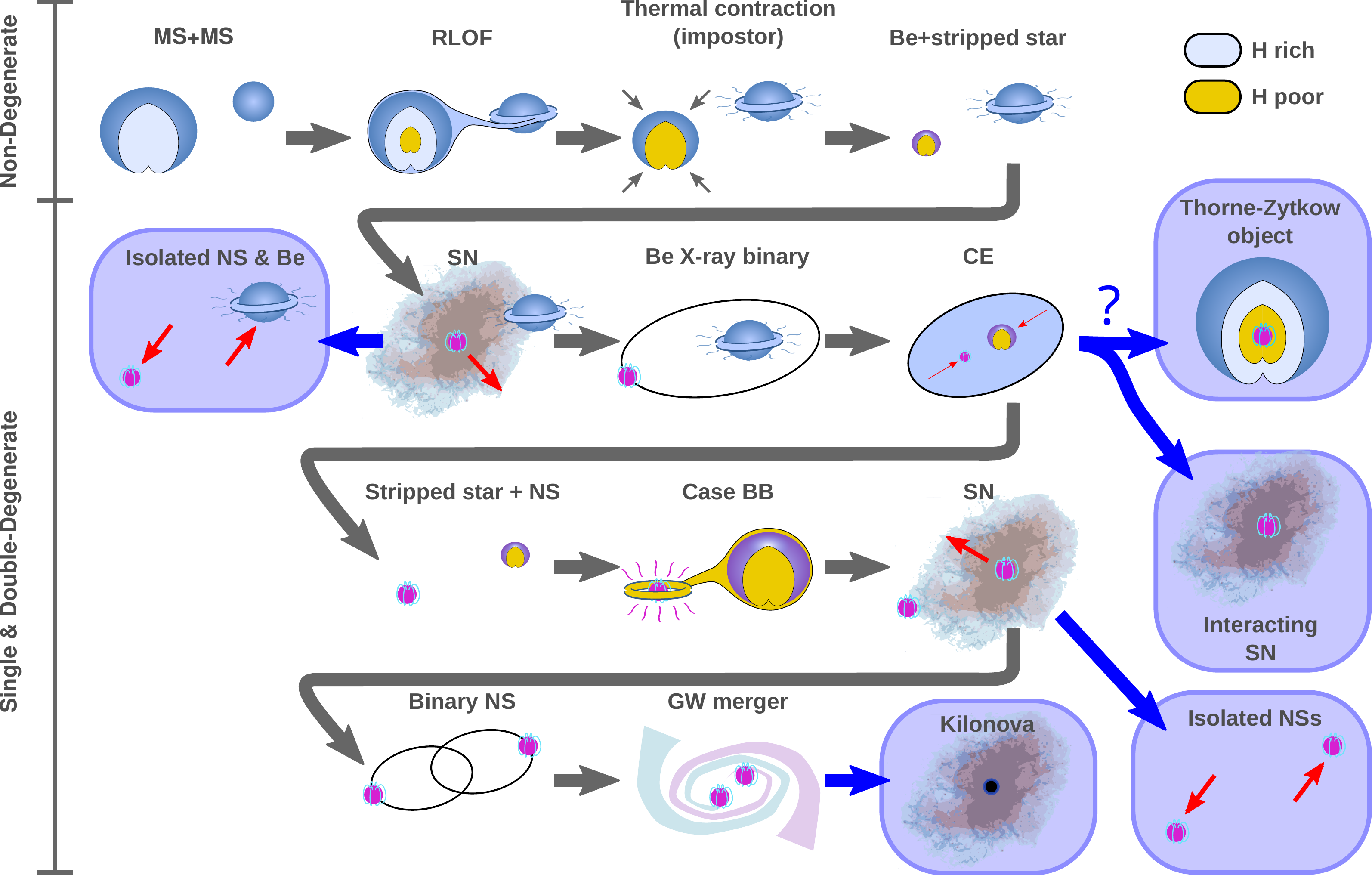}
    \caption{Evolution of a massive binary star from birth until the formation of a GW source, showing various intermediate post-interaction products. Phases in blue boxes represent end points for binary evolution.}
    \label{fig:post_interaction}
\end{figure}

An example of the evolution of a massive binary is schematically depicted in Figure\,\ref{fig:post_interaction}. It demonstrates the multitude of different intermediate stages of interaction products, both non-degenerate (described here) and single- or double-degenerate (described in Section\,\ref{sec:deg}, which have different observational characteristics. It also illustrates potential end points of binary evolution. Those depend on the parameters of the system and the type of interaction that occurs. Figure\,\ref{fig:HR_stripped} shows an observational overview of the mass donors in different mass regimes and evolutionary stages post-mass transfer in an HRD. We here include (candidate) objects detected by various techniques, which are described in the subsections below, and refrain from adding the mass gainers as often their parameters are not well constrained.

\subsection{Potential merger products}
Merger products are important in different aspects. As described in Sect.\,\ref{sec:binarity}, the BSG progenitor of SN1987A can be explained as a merger product in a binary context, but not with a single-star solution \citep[e.g.,][]{Podsi1990}. \citet{Schneider2015} discuss how rejuvenation in mergers can shape the mass function of stellar populations, while \citet{WangC2022} proposed that the blue component of the split MSs observed in young star clusters are formed by pre-MS mergers.

A handful of massive stars are potential stellar merger products. A famous example is the prototypical LBV $\eta$\,Car, which is currently in an eccentric binary system. Its giant eruption in the 19th century was proposed to be produced by the merging of the inner binary in a previous triple system \citep[][]{PortegiesZwart2016, Smith2018, Hirai+2021}. If other LBVs could be the product of a stellar merger remains to be constrained. Another proposed merger product is the slowly-rotating early B star $\tau$\,Sco \citep[spectral type B0.2V][]{Keszthelyi2021}. Its spectrum shows a strong nitrogen excess \citep[e.g.,][]{Martins2012}, and the star is associated with a large-scale complex magnetic field \citep[e.g.,][]{Petit2013}. \citet{Schneider+2019} reproduced the observed properties of  $\tau$\,Sco by a magnetohydrodynamic model of two merging MS stars, including its younger age in comparison to its parent association, the Upper Sco region. Based on a similar argument, \citet{Gies2022} interpreted the galactic runaway HD\,93521 as merger product: it appears younger than the time it would have taken to travel from the galactic disk to its current location. Interestingly, their rotational properties are quite different: while HD\,93521 is a rapid rotator, $\tau$\,Sco is an extremely slowly rotating star \citep{Nieva2014}

\begin{figure}
    \centering
    \includegraphics[width=\columnwidth]{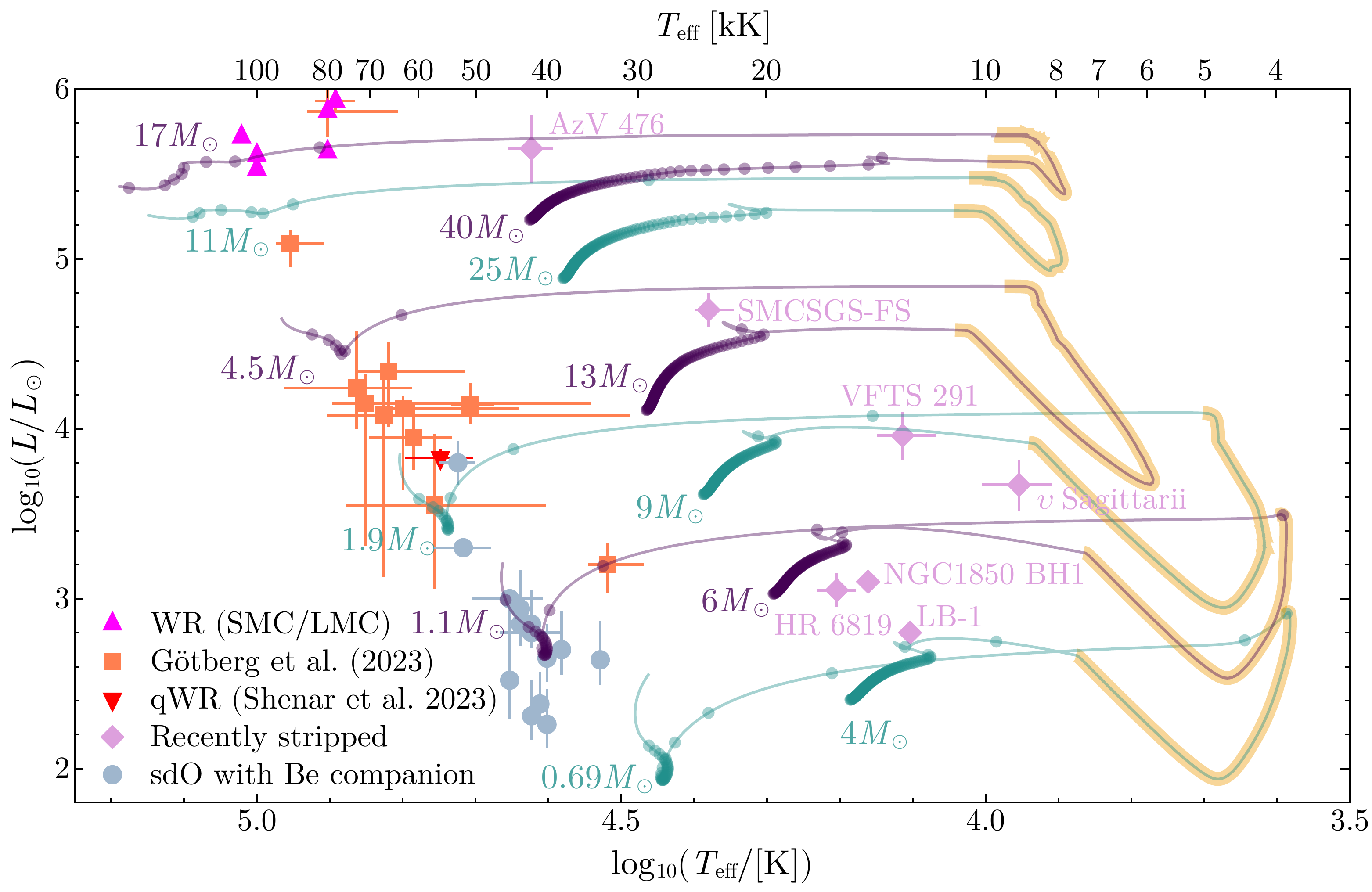}
    \caption{Observational sample of stripped stars and post-interaction binaries. The HRD shows the relation between bloated, recently stripped stars (light pink, see text for references) and their contracted successors \citep[][and references therein]{WangL2021}. It further shows that the mass gap between sdO stars (grey) and WR stars \citep[pink,][]{Shenar2016, Shenar2018, Shenar2020_WR} was recently filled with intermediate-mass stripped stars \citep[orange,][]{Gotberg2023}. Evolutionary tracks for donor stars of different masses undergoing case B mass transfer are shown, indicating their zero-age MS mass as well as their mass after stripping. Dots in the tracks are placed at intervals equal to 1\% of the MS lifetime. See text for details and references.}
    \label{fig:HR_stripped}
\end{figure}

Another proposed merger product is HD\,45166, which was previously denoted quasi-WR due to its spectral appearance \citep[it has similar, but narrower emission lines than normal WR stars, see e.g.,][and references therein]{Steiner2005}. It was initially interpreted as first example of an intermediate-mass stripped star bridging the mass gap between sdOBs and WRs (see Section \ref{subsec:stripped}), but was recently reported to exhibit a strong magnetic field of 43\,k$G$. This makes it a potential progenitor of a magnetar, a highly magnetized neutron star. To explain its properties, a merger of a hydrogen-rich star with a stripped star was proposed that expelled most of the H-rich envelope \citep{Shenar+2023}. Given the proposed involvement of a stripped star, we include it in Figure\,\ref{fig:HR_stripped}.

\subsection{Observations of stripped stars}\label{subsec:stripped}

It was long proposed that envelope stripping in cWRs occurs either because of their strong stellar winds \citep[e.g.,][]{Graefner2011} or due to mass transfer in a binary system \citep[e.g.,][]{Paczynski1967, Vanbeveren+1998}. It remains uncertain which of the two channels is the dominant one, which most likely also varies as a function of metallicity \citep[][]{Shenar2020_WR}. The viability of binary stripping is, however, demonstrated by the detection of lower-mass WR stars whose mass-loss rates are too low to strip their envelope \citep{Schootemeijer2018_WR}, in particular with rapidly-rotating companions \citep[e.g.,][]{Shenar2016, Shenar2019}. While cWRs have current masses above $\sim$10\,$\mathrm{M_\odot}$ and launch strong, optically thick stellar winds, sdOB stars are their equivalent at the low-mass end \citep[see][for a review]{Heber2009}. These core-helium burning objects with masses $\lesssim 1 \mathrm{M_\odot}$ are thought to be stripped stars and are found in binaries with white-dwarf, low-mass MS or OBe star companions \citep[e.g.,][and see Section\,\ref{subsec:gainer}]{Schaffenroth2022}. 

Stripped stars with masses in between have only recently been reported based on UV photometry and followed up with spectroscopy \citep{Drout2023, Gotberg2023}, bridging sdOs with WRs and thus filling a long-standing gap in our understanding of post-interaction binaries. In particular, their current masses were estimated to be between 1 and 8\,$\mathrm{M_\odot}$, putting them in the intermediate mass range between WR and sdOB stars. They show a variety of composite spectra with different amounts of contribution from potential companions whose nature remains to be constrained. Furthermore, all systems with multiple epochs of observations showed significant RV variations indicative of the presence of a companion.

\subsection{Observational constraints on mass gainers}\label{subsec:gainer}
Mass gainers from RLOF are expected to be rapidly rotating because of the accreted angular momentum (how rapidly still remains an open question, see Section 3). Based on a rapid population-synthesis calculation, 
\citet{deMink+2013} proposed that the rotational velocity distribution of massive stars is strongly shaped by binary interactions.

Indeed, the observed rotational velocity distribution of single OB stars in the 30 Dor region is bimodal, with a majority of stars rotating with velocities around 100\,km\,s${-1}$ and a tail of high velocities, interpreted as signature of previous interactions \citep{RamirezAgudelo2013, Dufton2013}. A similar signature was reported for the young SMC cluster NGC\,346 \citep{Dufton2019}. Focusing on the O-type primaries of detected binaries, \citet{RamirezAgudelo2015} further found a lack of very rapidly rotating primaries compared to single O-type stars, implying that prior interaction is required to produce such rapid rotators. Similar results were also reported for galactic O-type stars in the IACOB survey \citep{Holgado2022, Britavskiy2023}. The observed rotational velocities of stars in the slightly older cluster NGC~330 were interpreted as shaped by previous binary interactions \citep{Bodensteiner2023}.

If the stripped primary subsequently explodes as a SN, the system may be disrupted, potentially creating a rapidly-rotating runaway or walkaway star \citep[e.g.,][]{Blaauw1961, Renzo2019}. The galactic runaway star $\zeta$ Oph was reported as such \citep[with a space velocity of 30\,km\,s$^{-1}$, see e.g.,][]{Renzo2021}. \citet{Britavskiy2023} found that almost 65\% of apparently single fast-rotating O stars in IACOB are runaways. Investigating runaways in the VFTS, \citet{Sana2022} interpreted a population of rapidly rotating but slowly moving stars as results of binary ejections, in contrast to slowly rotating but rapidly moving stars interpreted as ejections by dynamical processes. Also the fraction of rapidly rotating OBe runaways matches binary population synthesis calculations \citep{Boubert2018}.  


Classical OBe stars, which are in general rapid rotators, were proposed to be interaction products (see Section\,\ref{sec:binarity}). Theoretical models agree on the feasibility of forming OBe stars according to this channel \citep[e.g.,][]{vanBever1997, Shao2014, Hastings2021}, but the predicted number of OBe stars formed by this channel varies from a few to basically 100\%. The numbers depends strongly on model assumptions and uncertain interaction physics, such as the mass-transfer efficiency or the reaction of the accretor. Another remaining open question is how close to critical rotation OBe stars are \citep{Rivinius2013}. If the binary channel dominates Be star formation, their multiplicity properties would be fundamentally different from 'normal' OB stars. Firstly, there should be no Be+MS binaries (unless they were with a third companion in an initial triple system), and secondly their companions should be stripped stars or compact objects. In some cases, a massive enough companion might have exploded, potentially disrupting the system.

Several well-known Be stars were proposed to be in binaries with an evolved companion. One example is the first Be star ever described, $\gamma$\,Cas, which is in a long-period binary system \citep[e.g.,][]{Harmanec2000}\footnote{$\gamma$ Cas also to shows hard, moderately strong X-ray emission \citep{Mason1976} making it the prototype of the so-called class of $\gamma$-Cas stars \citep[see e.g.,][for a recent compilation]{Naze2022}.}. Despite its brightness and various observational campaigns with different techniques, it remains debated if the companion is a white dwarf, a helium star or a neutron star \citep[e.g.][and references therein]{Langer2020gam_Cas}.

The detection of stripped sdOB companions to Be stars further matches expectations of the binary channel. Given the temperatures and radii of those stars, they are faint in the optical and mostly detected in the UV \citep[e.g.,][]{Gies1998, Peters2013, WangL2018, WangL2021, WangL2023}. In a handful of cases, there is a direct signature of the sdO star in the optical spectrum \citep[e.g. in $\phi$\,Per or FY CMa,][]{Poeckert1981, Rivinius2004}, or an observable change in the Be disk induced by the presence of the hot companion can be seen \citep[e.g. in $o$\,Pup,][]{Koubsky2012}. A handful of such systems were also detected through interferometry \citep[e.g., the first Be+sdB system $\kappa$ Dra,][]{Klement2022}.

\subsection{Thermally contracting stripped stars}
Recently, a new type of OBe binary was reported in a brief evolutionary phase in between mass transfer and the OBe+sdOB phase, in which the stripped star has not yet contracted and is still similarly bright in the optical as the mass accretor, the Be star (see Figure\,\ref{fig:post_interaction}). The first such systems reported were LB-1 \citep[e.g.,][]{Irrgang2020, Shenar2020_LB1} and HR\,6819 \citep[e.g.,][]{Bodensteiner2020_HR6819, ElBadry2021_HR6819}. Initially they were reported as binary or triple system hosting a BH \citep{Liu2019, Rivinius2020}, mainly because of two observational characteristics: firstly, in contrast to sdOB systems, the optical spectrum is dominated by the narrow-lined stripped star showing large RV amplitudes indicative of a high mass ratio \citep[in LB-1, the Be companion could only be revealed by spectral disentangling,][]{Shenar2020_LB1}.  Secondly, assuming the mass of the narrow-lined star, which appears like a 'normal' B-type star, from a comparison to single-star evolutionary tracks leads to a very high mass of the 'unseen' object. Given the proximity and brightness of HR~6819, the stripped star and Be companion with its disk could be resolved interferometrically \citep{Frost2022}. It was shown that the spectroscopic (and actual) masses of the stripped stars in LB-1 and HR~6819 are only $\lesssim 1\,\mathrm{M}_{\odot}$, and their companions are rapidly rotating Be stars with masses around 6\,$\mathrm{M}_{\odot}$. 

Other systems were thereafter detected with a similar signature but in which the mass gainer currently shows no emission lines, for example the still uncertain, highly debated case of NGC\,1850\,BH1 \citep{Saracino2022, Saracino2023, ElBadry2022_NGC1850}. Additional systems interpreted to have more massive stripped, bloated companions are VFTS\,291 \citep{Villasenor2023}, SMCSGS-FS\,69 \citep{Ramachandran2023} and AzV\,476 \citep{Pauli2022}. 

These thermally contracting systems provide a critical snapshot of a binary systems right after the interaction occurred (see Figure\,\ref{fig:HR_stripped}).
A common feature is their mass ratios ($q=5$ for LB-1, and $q\sim15$ for HR\,6819), which were reported to require conservative mass transfer. While those systems are in a short evolutionary phase, they are more easily detectable with common observing techniques than their subsequent, longer-lived evolutionary stage (OB(e) + sdOB/WR systems) because of the higher optical brightness of the stripped star. \citet{ElBadry2021_HR6819} also showed that the luminosity of the stripped stars is not purely powered by contraction, but also by shell-helium burning. This extends the lifetime of the contraction phase beyond a simple thermal timescale, making it more likely to observe binaries in this phase.

Another stripped star binary in a later evolutionary stage is the helium supergiant $\upsilon$\,Sagittarius \citep{Gilkis2023}, which was reported to be in a currently interacting binary system during a second phase of mass-transfer. It was proposed that the more luminous primary is stripped of the remainder of its hydrogen envelope, while the accretor is a rapidly-rotating B-type star. The stripped primary of $\upsilon$\,Sag is also included in Figure\,\ref{fig:HRD_binary_stats} (the evolutionary tracks do not include this second phase of mass transfer).

The aforementioned observed systems form an evolutionary sequence (see Figure\,\ref{fig:HR_stripped}) of recently stripped stars with OBe companions that later contract and appear as sdOB+OBe binaries when they are core-helium burning. Spectroscopically, the recently stripped stars in a contraction phase look like normal B-type MS stars. This is in contrast to expected surface abundances of stripped stars, which are thought to be dominated by helium \citep{Gotberg2017, Schurmann2022}. A potential explanation for the hydrogen-rich surface is that they have re-accreted hydrogen-rich material from the Be star decretion disk \citep{Bodensteiner2020_HR6819}. So far, a common characteristic of mass gainers is their rapid rotation. The fact that some of the potential mass-gainer companions do not show emission lines could simply be related to the transient nature of the Be phenomenon \citep[e.g.,][]{Rivinius2013}. The detection of additional such systems will allow to better constrain mass-transfer physics, the response of the accretor, and how close to the critical velocity it is spun up.

\section{SINGLE AND DOUBLE DEGENERATE BINARIES}\label{sec:deg}

\begin{figure}
    \centering
    \includegraphics[width=\columnwidth]{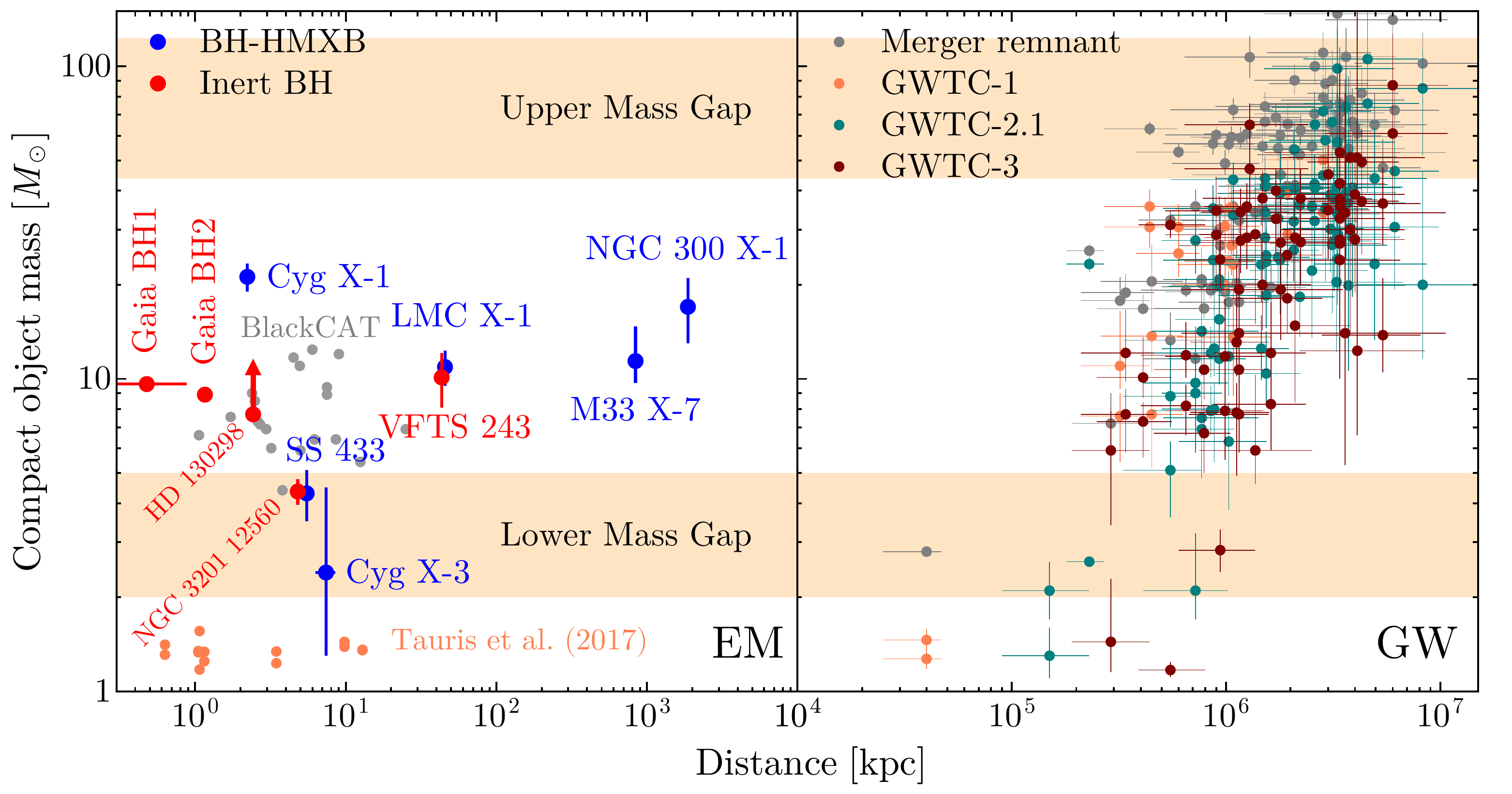}
    \caption{(left) Masses and distances to BH-HMXBs with measured dynamical masses, as well as detected inert BHs. Also shown are the masses of low mass X-ray transients with dynamical mass estimates from the BlackCAT catalogue \citep{Corral-Santana+2016} and the masses of binary neutron stars collected by \citet{Tauris+2017}. (right) masses of merging compact objects detected through GW emission. A representative band between $2-5\;\mathrm{M_\odot}$ is shown for the potential mass gap between NSs and BHs, while the upper mass gap due to pair-instability SNe is taken from \citep{Marchant+2019}.}
    \label{fig:CO_masses}
\end{figure}

Once one of the components in a massive binary undergoes core-collapse, it can produce a neutron star (NS) or black hole (BH). The mapping between pre-explosion properties and the type and mass of the compact object depends on the physics of the supernova process (or lack thereof). A common method used to assess the post-explosion outcome is the compactness parameter, taken as a ratio between the mass and the radius at a specific mass coordinate \citep{OconnorOtt2011}. Stars with higher compactness in their central regions are expected to collapse into a BH. Stellar evolution calculations done until core-collapse consistently show that, based on their compactness as well as other metrics, the boundary between NSs and BHs is not a simple mass threshold, but rather consists of ``islands'' of explodability \citep{Ugliano+2012,SukhboldWoosley2014, Sukhbold+2016}. \citealt{Schneider+2023} suggests that these variations in explodability lead to peaks in the mass distribution of BHs observed in gravitational wave sources. 

Whether or not the compact object remains bound to its companion depends on the total mass ejected and the kick imparted onto it. Observations of the proper motion of isolated NSs have been used to assess the distribution of kick velocities \citep{Hobbs+2005}, and are often utilized in both rapid and detailed evolutionary calculations. Considering only isolated pulsars, however, is biased towards the strongest kicks (eg. \citealt{Odoherty2023}), which can lead to an overestimate of unbound post-SN systems in population synthesis. Isolated BHs can be detected through lensing of their background stars, with a recent first detection (although data could also support a NS as the lens, \citealt{Lam+2022, Sahu+2022}). Data on isolated BHs is thus insufficient to estimate BH kick distributions, and instead BHs in X-ray binaries have been the standard method to determine BH kicks (e.g., \citealt{Atri+2019}), showing support for both strong ($>50\;\mathrm{km\;s^{-1}}$) and weak kicks at birth. One weakness of the methods that use low mass X-ray binaries to determine kicks is that they can only constrain the velocity imparted on the system at birth, and not the magnitude of the kick imparted onto the BH itself.

If the binary remains bound after the formation of the first or second compact object, and if the end result is not a binary BH, we can use electromagnetic observations to infer its orbital parameters, serving as a laboratory of the SN process. A detailed understanding of the intrinsic population of single-degenerate binaries is also crucial as an anchor point to constrain evolutionary models. \citet{Tauris+2017} recently provided an extensive review on binary NSs, including their formation and intermediate stages. Thus, we only provide a brief overview of NSs in massive binaries and focus our discussion on the properties of BH-high mass X-ray binaries (BH-HMXBs) as well the recently identified category of inert BHs (corresponding to BH binaries wide enough that little mass is transferred and no accretion disk is formed). A compilation of masses and distances of NSs and BHs in massive binaries is shown in Figure \ref{fig:CO_masses}, including also GW observations. The detection of BHs through GWs has significantly outnumbered the electromagnetic sources for which we have mass constraints, but the population of inert BHs could grow by orders of magnitude in the coming decade (see Sect.\,\ref{sec:inert}).

\subsection{Neutron stars in massive binaries}\label{sec:nss}
Be X-ray binaries are composed of a compact object that is fed from the decretion disk of its companion Be star (\citealt{Negueruela1998,OkazakiNegueruela2001}, see \citealt{Reig2011} for a review). Only one system has been claimed to contain a BH rather than a NS \citep{Casares+2014}, but its status is currently contested with the compact object potentially being a stripped helium star instead \citep{Rivinius+2022, Janssens+2023}. Assuming that a previous mass-transfer phase circularised the orbit before the SN, the eccentricity distribution of Be X-ray binaries is indicative of a subpopulation that underwent small kicks at NS formation \citep{Pfahl+2002}. One explanation for this is the occurrence of electron capture SNe rather than iron-core collapse SNe \citep{Podsiadlowski+2004}. The masses of Be stars in X-ray binaries have been used to argue that they underwent an efficient mass-transfer process \citep{Vinciguerra+2020}, but masses for Be stars are typically inferred from their spectral type, which carries significant uncertainty.

In OB+NS systems, after the OB star finishes its MS evolution, it will expand and fill its own Roche lobe (see Fig.\,\ref{fig:post_interaction}). Owing to the large mass ratio, this most likely results is CE evolution, forming a compact stripped star+NS system, or instead merging. The outcome of a compact object merging with a stellar companion is uncertain, with possible outcomes being a stable Thorne-Zytkow object (\citealt{ThorneZytkow1975}, see \citealt{Farmer+2023} for a recent picture) or a SN explosion \citep{Chevalier2012, Metzger+2022}. Systems that survive CE and have stripped stars with masses $\lesssim 5\,\mathrm{M_\odot}$ can undergo an additional phase of mass transfer after core helium depletion, which is referred to as Case BB mass transfer\footnote{Consistent nomenclature would use the name case BC for this phase, as it happens after core helium depletion.} \citep{Tauris+2015}. This mass-transfer phase can leave an almost stripped helium core, producing SNe with $\lesssim 0.1\,\mathrm{M_\odot}$ of ejecta and potentially small kick velocities \citep{Moriya+2017}. 

If the binary remains bound after the second SN, a double degenerate binary is produced. Neutron stars in binaries that are detected as radio pulsars provide an accurate probe of their radial motion through measurements of variations in the time of arrival of pulses. It was through timing measurement of the first radio pulsar \citep{HulseTaylor1975} that gravitational waves were first measured indirectly (\citealt{Taylor+1979}, see \citealt{Weisberg+2010} for results with over three decades of timing data). Another remarkable case is the double pulsar \citep{Kramer+2006, Kramer+2021}, where the detection of multiple post-Newtonian effects in the orbit allows the measurement of the individual pulsar masses to better than one part in ten thousand. Currently there are over 20 binary NSs detected (see \citealt{Bernadich+2023} for a recent collection), and the sample is expected to grow by up to an order of magnitude with the advent of the Square Kilometer Array (SKA, \citealt{Keane+2015}). Even though binary NSs are usually discussed in the context of isolated binary evolution, it is important to consider alternative formation scenarios. \citet{AndrewsMandel2019} have suggested that the current sample contains a sub-population that is inconsistent with binary evolution and could have formed through dynamical processes.

\subsection{Black-hole high-mass X-ray binaries}
The X-ray flux of young star-forming galaxies without active galactic nuclei is dominated by HMXBs. The \textit{Chandra} telescope has played a pivotal role in their study, allowing for the observation of X-ray binaries in galaxies beyond the local group (see \citealt{Gilfanov+2022} for a recent review). The brightest of these sources, with luminosities exceeding $10^{39}\;\mathrm{erg\;s^{1}}$, are referred to as ultraluminous X-ray sources (ULXs, \citealt{LongvanSpeybroeck1983}, see \citealt{Kaaret+2017} for a recent review). ULXs exceed the Eddington limit for a $10\,\mathrm{M_\odot}$ BH, in some cases by over an order of magnitude, and as such they have been suggested to hold intermediate-mass BHs. This needs not be the case if there is significant beaming of radiation, which \citet{King+2001} argues would be the case for rapid thermal-timescale mass transfer phases onto compact objects. Even more contrary to the idea that ULXs host intermediate mass BHs, the discovery of X-ray pulsations in some systems allowed a clear identification of their compact object accretors to be NSs \citep{Bachetti+2014}. A clear identification of a BH accretor in a ULX is more elusive, but within our own galaxy it is suggested that the microquasar SS 433 (which hosts a BH) would appear as a ULX if observed from a different angle \citep{Begelman+2006}.

For extragalactic X-ray binaries, spectroscopic follow-up that would allow the measurement of RVs and an estimate on the compact-object mass is challenging. Even when spectral features associated to the donor are detected, observed RV variations might not follow the orbital motion. Such is the case of IC10 X-1, which contains a WR star orbiting a compact object with a period of 35 hours. RV variations of the WR star indicated that IC10 X-1 hosts the most massive stellar-mass BH known \citep{Prestwich+2007,SilvermanFilippenko2008}. However, \citet{Laycock+2015} showed that the phase of the RV variability was inconsistent with eclipses in the system and as such did not trace the orbital motion of the WR star, making the nature of the compact object unclear. Important adjustments have also been made to the measured masses of the BHs in Cyg X-1 and M33 X-7. A more accurate distance determination to Cyg X-1, placing it further away from us, has increased the mass estimate of its donor by $\sim 30\%$ leading to a corresponding increase on its estimated BH mass \citep{Miller-Jones2021}. For M33 X-7, detailed spectroscopic modeling of its donor star has instead lowered the donor mass estimate by $\sim 30\%$, leading to a lower mass estimate for its BH \citep{Ramachandran+2022}. 

In the presence of an accretion disk around the BH, it is also possible to constrain its dimensionless spin (see \citealt{Belczynski+2021} for an overview of methods, as well as caveats). Table \ref{tab:inertBH} lists the three known BH-HMXBs with spin estimates, all of which are high. For M33 X-7, \citet{Ramachandran+2022} argued that the lowered mass estimate on the BH would lower its measured spin from $0.84\pm{0.05}$ \citep{Liu+2010} down to $\sim 0.6$. For both Cyg X-1 and LMC X-1, the spin of the BH is near critical rotation. The origin of the high spin of these BH is unclear. It has been suggested that the source of the angular momentum in the BH progenitor came from tidal synchronization during an earlier stage of mass transfer \citep{Valsecchi+2010, Qin+2018} or from a failed SN where some mass expanded and gained angular momentum from the orbit before falling back into the newly formed BH \citep{Batta+2017}. Understanding the origin of spin in BH-HMXBs is crucial to understanding their potential link to GW sources, which are mostly observed to have low spins (see the discussion by \citealt{FishbachKalogera2022}). If BH-HMXBs inherit their spins from tidal coupling, longer-period systems could potentially exhibit lower spins, but it is unclear at which orbital period the BH would become X-ray inactive.

\begin{table}
\caption{Properties of BH-HMXBs and inert BHs, giving the distance d, orbital period P, mass of the compact object M$_\mathrm{CO}$, mass of the companion star M$_\mathrm{comp}$, and spin. Cyg X-3 could potentially have a neutron star instead of a BH. The error on the orbital period for all systems is below $1\%$.}\label{tab:inertBH}
\begin{center}
\begin{tabular}{l l l l l l} \hline
Name & d & P & M$_\mathrm{CO}$ & M$_\mathrm{comp}$ & BH spin \\
 & [kpc] & [days] & [$\mathrm{M_\odot}$] & [$\mathrm{M_\odot}$] & \\ \hline
\multicolumn{6}{c}{BH-HMXBs} \\ \hline
Cyg X-1 & 2.22$^{+0.17}_{-0.18}$ & $5.60$ & $21.2 \pm 2.2$ & $40.6^{+7.7}_{-7.1}$ & $>0.9985$ \\ 
Cyg X-3 & $7.4 \pm 1.1$ & $0.200$ & $2.4 \pm 2.1$ & $10.3^{+3.9}_{-2.8}$ & - \\ 
NGC 300 X-1 & $1870^{+115}_{-108}$ & $1.37$ & $17 \pm 4$ & $26^{+7}_{-5}$ & - \\ 
M33 X-7 & $840^{+27}_{-28}$ & $3.45$ & $11.4^{+3.3}_{-1.7}$ & $38^{+22}_{-10}$ & $0.84\pm 0.05$ \\ 
LMC X-1 & $45.59 \pm 0.09$ & $3.91$ & $10.91 \pm 1.41$ & $31.79 \pm 3.48$ & $0.92^{+0.05}_{-0.07}$ \\ 
SS 433 & $5.5 \pm 2.2$ & 13.1 & $4.3 \pm 0.8$ & $12.3 \pm 3.3$ & - \\ \hline 
 \multicolumn{6}{c}{inert BHs} \\ \hline
 VFTS 243 & $45.59 \pm 0.09$ & $ 10.4$ & $10.1\pm2.0$ & $25.0 \pm 2.3$ & - \\
 HD 130298 & $2.425^{+0.08053}_{-0.07507}$ & $14.6$ & $>7.7\pm1.5$ & $24.2\pm 3.8$ & - \\ 
 Gaia BH1 & $0.47 \pm 0.4$ & $186$ & $9.62 \pm 0.18$ & $0.93 \pm 0.5$ & -  \\  
 Gaia BH2 & $1.16 \pm 0.02$ & $128$ & $8.94 \pm 0.34$ & $1.07 \pm 0.19$ & -  \\ 
NGC\,3201\,12560 & 4.8 & $167$ & $4.36 \pm 0.41$ & $0.81 \pm 0.05$ & -  \\ 
\hline
\end{tabular}
\end{center}
\begin{tabnote}
References: Cyg\,X-1 \citep{Miller-Jones2021,Mahy2022_BH}, 
Cyg\,X3 \citep{Zdziarski2013, McCollough2016, Singh2002},
NGC\,330 X-1 \citep{Binder2021, Rizzi2006, Crowther2010}, 
M33\,X-7 \citep{Ramachandran2023, Gieren2013, Pietsch2006, Liu+2010}, 
LMC\,X-1 \citep{Orosz2009, Gou+2009, Pietrzynski+2019},
SS\,433 \citep{Hillwig2008, Blundell2004}, 
VFTS\,243 \citep{Shenar2022, Pietrzynski+2019}, 
HD\,130298 \citep{Mahy2022_BH},
Gaia\,BH1 \citep{ElBadry2023_BH1},
Gaia\,BH2 \citep{ElBadry2023_BH2},
NGC\,3201\,12560 \citep{Giesers2018}.
\end{tabnote}
 \end{table}

To get detectable X-ray fluxes in BH binaries,  the BH not only has to accrete sufficient mass, but also an accretion disk must form \citep{ShapiroLightman1976}. Taking this into account, \citet{Vanbeveren+2020} suggested that if the known WR+O binary systems in the solar neighborhood would evolve to become BH+O binaries, there should be over 100 BH-HMXBs within a few kiloparsec of the Sun. A solution to that discrepancy would require most WR+O binaries to undergo a SNe and form a NS, or form a BH with a strong kick that unbinds the system. \citep{Sen+2021} improved upon the model for disk formation of \citet{Vanbeveren+2020}, and showed that owing to the fast winds of OB stars, only the closest OB+BH binaries would be observed as X-ray sources. Similarly, \citet{HiraiMandel2021} suggested that tidal deformation in near-Roche-filling binaries leads to slow wind outflows through the vicinity of the L1 point, which can efficiently be captured by the BH and produce an accretion disk. If only near-Roche-filling systems can form discs around BHs, then WR+O star systems can still primarily evolve towards BH+O star binaries without being in conflict with the low number of observed BH-HMXBs. The vast majority of BHs with massive companions would then be inactive in X-rays.

\subsection{Inert black holes}\label{sec:inert}
Recent work on the proposed population of long-period BHs which do not form accretion disks has referred to them as ``quiescent'' or ``dormant''. However, this denomination can be confused with nomenclature used for X-ray active binaries, which can undergo periods of X-ray inactivity while still containing an accretion disk. We instead adopt the term ``inert'' for these BHs, to indicate that other than their gravitational influence on their companions, we do not expect them to become X-ray active in the near future. Using detailed population synthesis calculations, \citet{Langer+2020} has suggested that $\sim 3\%$ of OB binaries contain BH companions, and almost all would be inert.

There are three main methods proposed to identify inert BHs in binaries. In low-mass BH binaries, ellipsoidal variability near Roche filling for the donor star can be identified through photometric measurements (eg. \citealt{Gomel2021}), although with a massive star donor it is expected that wind mass transfer would already make it active at this stage. Alternatively, the reflex motion of the companion star can be determined either through astrometry and/or by measuring its RV through spectroscopy. Currently it is only via spectroscopy and astrometry that confident detections have been made.

In the context of the \textit{Gaia} mission, multiple studies pointed out the possibility to detect hundreds to thousands of inert galactic BHs in binaries \citep{Breivik+2017,MashianLoeb2017,Yamaguchi+2018,Yalinewich+2018,Wiktorowicz+2019,Janssens+2022}. Although this was expected to happen with the third data release of Gaia, the stringent criteria that were placed on the data in order to release orbital solutions excluded almost all massive stars \citep{Janssens+2023a}. Current discoveries have instead been made in the low-mass regime, with \citep{ElBadry2023_BH1,ElBadry2023_BH2} reporting two confident detections of $\sim 9\mathrm{M_\odot}$ BHs orbiting $\sim 1 \mathrm{M_\odot}$ stars.

Spectroscopic measurements have allowed the first identification of inert BHs with massive companions. As part of the TMBM survey, \citet{Shenar2022} followed up on the 51 apparent SB1 systems using spectral disentangling techniques to exclude the presence of a second luminous component. This allowed them to discover VFTS 243 \citep{Shenar2022b}, a binary consisting of a $25\pm2.3 \mathrm{M_\odot}$ O star and a $10\pm 2 \mathrm{M_\odot}$ BH companion in a near-circular orbit ($e=0.017\pm0.012$). Similarly, \citet{Mahy2022_BH} studied a sample of 32 Galactic SB1 stars and identified HD 130298 as an O+BH system, with masses similar to those of VFTS 243 but an eccentricity $e=0.457\pm0.007$. Contrary to BH-HMXBs, where large filling factors mean tidal forces are expected to lead to rapid circularisation, the eccentricity in inert BHs provides a direct constrain on the SN kick at their formation. \cite{Shenar2022b} has argued that the BH in VFTS 243 most likely formed with a weak kick and $<1 \mathrm{M_\odot}$ of ejecta. Spectroscopic observations have also allowed for the detection of an inert BH with a low mass companion \citep{Giesers2018} as well as two additional candidates \citep{Giesers+2019}.

\section{GRAVITATIONAL WAVE SOURCES}\label{sec:GW}

Gravitational wave (GW) observations are emerging as a very fertile ground to study the evolution of massive and very massive stars. It is also very rapidly evolving, and as such, our objective here is not to provide a thorough review of the current state of this field. Rather, we aim to discuss the prospects of future GW observations as well as the binary processes that are thought to contribute to the observed sample. 

\subsection{Observations of gravitational wave sources}
Second generation GW observatories (including advanced LIGO, advanced Virgo and KAGRA) are currently on their third observing round \citep{GW_observing_prospects}. The teams that develop each of these instruments are assembled into the LIGO-Virgo-KAGRA collaboration and up to the moment of writing have released almost a hundred detections, which are compiled in the Gravitational Wave Transient Catalogues (GWTCs, \citep{GWTC-3}). For any given compact-object merger, the three best-constrained quantities are the chirp mass, the mass ratio\footnote{Mass ratios in GW observations are normally defined as the ratio between the least and the most massive component, and as such they are smaller than unity.} and the effective spin, defined respectively as:

\begin{eqnarray}
    \mathcal{M} = \frac{(m_1 m_2)^{3/5}}{(m_1+m_2)^{1/5}}, \quad q=\frac{m_2}{m_1}, \quad \chi_\mathrm{eff}=\frac{m_1\chi_1 + m_2 \chi_2}{m_1+m2},
\end{eqnarray}
where $\chi_1$ and $\chi_2$ are the components of the dimensionless spin of each compact object that are aligned with the orbital angular momentum. In practice, the frequency evolution of the source provides a measurement of the redshifted chirp mass $(1+z)\mathcal{M}$, and to determine $\mathcal{M}$ one needs to make an assumption on the cosmology. Cases where an associated electromagnetic transient is detected are of particular importance, as a separate redshift measurement allows for the use of GW sources as standard candles, as was the case for GW170817 \citep{GW170817}. The chirp mass is directly derived by the frequency evolution of the binary, so for binaries that undergo many cycles in-band it can be accurately determined. For instance, the error on the chirp mass for GW170817 was smaller than 0.5\%. Most measurements of $q$ and $\chi_\mathrm{eff}$ show a significant degeneracy with current detector sensitivity, which can make it difficult to differentiate the nature of the merging compact objects \citep{Hannam+2013}. In some sources additional information has been obtained by measuring precession \citep{Hannam+2022,Varma+2022} as well as ringdown frequencies \citep{Abbott+2021GRtests}.

The long-term prospects for GW astrophysics are certainly exciting. As the strain that is measured is proportional to the inverse of the luminosity distance (rather than an inverse square dependence as is the case for electromagnetic waves), detector improvements significantly increase the volume to which it is sensitive. This will possibly result in $\sim10^4$ detections by the end of the decade \citep{Baibhav+2019}. In their expected sensitivity, third generation detectors will probe the Universe down to the redshift of the formation of the first stars \citep{HallEvans2019}, making GW mergers the best characterized astrophysical population. Space-based missions such as the Laser Interferometer Space Antenna (LISA) will operate at lower frequencies, and allow for multi-band detections of stellar-mass merging binary BHs \citep{Sesana2016}.

Besides providing a probe into new astrophysical environments, two aspects of GW science make it very interesting to contrast observations against the theory of binary evolution. On the one hand, detector biases are very well understood, allowing for a transparent determination of intrinsic source properties. Owing to the limited amount of observations, at the moment most distribution properties are determined using parameterized models that contain expected features such as the lower and upper BH mass gaps \citep{GWTC3-pop}, but reliance on such models will lower as the number of detections increases. On the theoretical side, given a model that produces the rate of formation of double compact objects (including their masses, separations and spins), it is straightforward to provide expected rates of observation for specific detectors, as the time it takes for two compact objects to merge is easily calculable from general relativity \citep{Peters1964}.

\subsection{Binary formation channels of gravitational wave sources}
Although one could expect the natural formation scenario of a merging binary BH to involve the evolution of two stars in a binary system, a wide variety of processes that do not involve binary interaction are being considered (see \citealt{MandelFarmer2022} for an overview of some of the proposed channels). The relative contribution of different formation scenarios to the observed population is still uncertain, although comparison of predicted distributions to observations suggests that multiple processes contribute to it (e.g., \citealt{Zevin+2021}). One important thing to keep in mind is that evolutionary paths towards a merging compact object are a rare outcome of massive-star evolution. Currently measured merger rates at redshift zero, in units of $\mathrm{Gpc^{-3}\;yr{-1}}$, range between $10-1700$, $7.8-130$ and $16-61$ for NS+NS, NS+BH and BH+BH mergers \citep{GWTC3-pop}. In contrast, the rate for core-collapse SNe is on the order of $10^5\;\mathrm{Gpc^{-3}\;yr^{-1}}$. Independent of the formation process, the main challenge to form a GW source is that the resulting binary BH needs to remain in a very compact orbit. For a circular compact object binary with a given orbital period and chirp mass, the time it takes for GWs to produce a merger is \citep{Peters1964}:
\begin{eqnarray}
    t_\mathrm{delay} = 11.9\;\mathrm{Gyr}\;\left(\frac{P}{5\;\mathrm{[days]}}\right)^{8/3}\left(\frac{\mathcal{M}}{30 \mathrm{M_\odot}}\right)^{-5/3}.
\end{eqnarray}
This implies that massive binaries that evolve to form merging binary BHs necessarily interact during their lifetime, and they need not only to survive to form a compact object binary but also finish their evolution in a short-period orbit. 

The classical formation channel used to explain their formation is CE evolution. Shortly after \citet{Paczynski1976b} proposed CE as the formation scenario for cataclysmic variables, \citet{vandenHeuvel1976} suggested it could explain the formation of the Hulse-Taylor binary pulsar, while \citet{TutukovYungelson1993} argued it could also produce binary BHs (see \citealt{Bavera+2021} for a recent overview). An alternative scenario considers the case where both stars in a (near-)contact binary evolve chemically homogeneously \citep{MandelDeMink2016,Marchant+2016} and form a binary BH. Chemically homogeneous evolution can also lead to the formation of merging BH+NS systems \citep{Marchant+2017}. The third formation process that is commonly considered to form merging binary BHs is stable mass transfer \citep{vandenHeuvel2017}, where a short-period binary is formed through non-conservative mass transfer from a star onto the first formed BH in the system. \citet{Picco+2023} have suggested that stable mass transfer can also naturally produce NS+BH mergers, and that the mechanism is robust against uncertainties on the angular momentum budget of binary outflows. Current work suggests that population synthesis studies have systematically overestimated the contribution of the CE channel to the formation of merging binary BHs, owing to issues with adopted binding energies and criteria for unstable mass transfer \citep{Klencki+2021,Olejak+2021, Marchant+2021, Gallegos-Garcia+2021}. One channel that is not often discussed is that of pop III evolution, in which case the absence of metals is expected to lead to much smaller stellar radii, making it easier to form compact binaries by the end of their evolution \citep{Kinugawa+2014}.

A large number of studies have performed population synthesis (both using detailed and rapid codes) to compute the rates and distributions of compact object mergers. \citet{MandelBroekgaarden2022} have made a compilation of published rate predictions, not only restricted to binary evolution, illustrating how rates for individual channels not only have uncertainties exceeding an order of magnitude, but that predictions from different groups can also have order of magnitude discrepancies. This is not entirely surprising, since predicting merging compact objects requires an understanding of all phases of massive binary evolution coupled with the metallicity-dependent star-formation history. Even for individual systems, different evolutionary codes can produce wildly different outcomes both for single-star evolution \citep{Romagnolo+2023} as well as binary evolution \citep{Belczynski+2022}. Moving forward will require collaboration within different research groups to identify the sources of these modeling discrepancies, as well as to define the different characteristic features of the distribution of merging compact objects that remain invariant independent of physical uncertainties \citep{vanSon+2023}.

\section{CONCLUSIONS}\label{sec:concl}
In the last decades, the field of massive binary evolution has grown significantly in terms of interest and scientific advances. This was driven first and foremost by large-scale spectroscopic observations, high-quality continuous photometric monitoring and high-precision astrometric survey. A consensus has emerged that binary evolution dominates the lives of massive stars, providing exciting new evolutionary pathways but also complicating our theoretical description of stellar evolution.
In recent years, massive stars gained further visibility due to the detection of GWs, which opened a new window to constrain potential end products of massive star evolution. As a conclusion, we provide below summary points as well as future issues that we expect will shape the field in the coming decade.

\begin{summary}[SUMMARY POINTS]
\begin{enumerate}
\item A game changer in recent years have been large-scale surveys with well-understood biases, not only for electromagnetic observations in terms of spectroscopic, photometric or astrometric surveys, but also in terms of GW detections, for which detection biases are well constrained. Those have corroborated the finding that binary interactions dominate massive star evolution.
\item Observations tentatively indicate that the multiplicity properties of OB stars, both binary fractions as well as orbital parameter distributions, are universal across the probed metallicity environments of the Milky Way, the LMC and SMC.
\item Theoretical models of binary evolution rely on three distinct types of calculations, including 3D hydrodynamical codes, 1D stellar-evolution calculations and rapid semi-analytical approximations. Each of these play a critical role. 
\item Observational evidence is growing that a majority of rapidly rotating stars, in particular OBe stars, can be explained as accretors in previous binary interactions.
\item In recent years, our picture of post-interaction products has been expanded to include massive stripped stars (closing the previous gap between sdOs and WRs) as well as thermally contracting objects right after mass stripping. These provide a novel perspective onto the mass transfer process.
\item Inert BHs without accretion disks or detectable X-ray radiation have been identified in binary systems. These extend the known population of BHs to longer periods, and a much larger population will potentially be unveiled in the coming years.
\item The number of GW observations has gone well beyond that of compact objects with dynamical masses measured from electromagnetic radiation. They provide a new window into massive star evolution, but their formation processes remain unclear.
\end{enumerate}
\end{summary}

\begin{issues}[FUTURE ISSUES]
\begin{enumerate}
\item The multiplicity fractions and orbital parameters of key objects such as OBe stars, WR stars or BSGs, remain poorly understood. Constraining those can provide us with important new insights on their evolutionary status and connection. 
\item Stellar evolution models still rely on simplified assumptions regarding how conservative mass transfer is, and the dynamics of outflows from binary systems. Narrowing down this uncertainty will require careful studies that combine detailed hydrodynamical simulations with long-term evolutionary calculations.
\item Increased detections of inert black holes will provide a critical anchor point for binary evolution models. They will also give novel constraints on the supernova process and, in particular, potential kicks imparted onto the BH at birth. This will also require a clear understanding of the pre-SN binary properties, in particular the eccentricity and period distribution of WR+OB binaries.
\item 
Recent observations indicate that triple- and higher-order multiple systems are common, even more so for more massive stars. Those could be important probes of the star-formation process and play a crucial role by inducing binary interactions through dynamical processes.
\item It is critical to benchmark binary evolution models against each other, to clearly identify which physical assumptions lead to discrepant results.
\item It is also important to keep in mind that significant uncertainty exists still in massive single-star evolution, which translates directly into binary evolution. In particular, the radial evolution of stars near the Eddington limit is poorly understood.
\item New instrumentation on upcoming 40-m-class telescopes, such as HARMONI or MICADO at ESO's Extremely Large Telescope (ELT, first light expected before the end of the decade), will push our understanding of massive-star evolution to lower metallicity, for example by resolving individual stars in distant stellar populations.

\end{enumerate}
\end{issues}

\section*{DISCLOSURE STATEMENT}
The authors are not aware of any affiliations, memberships, funding, or financial holdings that might be perceived as affecting the objectivity of this review. 
%
\section*{ACKNOWLEDGMENTS}
The authors thank T. Shenar and A. Istrate for always insightful discussions and comments on the manuscript. PM acknowledges support from the FWO senior postdoctoral fellowship No. 12ZY523N.

%
\bibliographystyle{ar-style2}

\end{document}